\documentclass[journal,comsoc]{IEEEtran}
\usepackage[T1]{fontenc}% optional T1 font encoding
\usepackage{cite, color}
\usepackage[normalem]{ulem}
\usepackage{longtable}
\usepackage{float}
\usepackage[acronym,nonumberlist,nogroupskip,style=super]{glossaries}
\usepackage{array}
\usepackage{multirow}
\usepackage{booktabs}
\usepackage{multirow}
   \usepackage [pdftex]{graphicx}
   \DeclareGraphicsExtensions{.pdf,.jpg,.png,.eps}
\usepackage{amsmath}
\usepackage{inputenc}
\usepackage{amssymb}
\usepackage{enumitem}
\usepackage[cmintegrals]{newtxmath}
\usepackage{newtxtext}
\usepackage[export]{adjustbox}
\ifCLASSOPTIONcompsoc
  \usepackage[caption=false,font=normalsize,labelfont=sf,textfont=sf]{subfig}
\else
  \usepackage[caption=false,font=footnotesize]{subfig}
\fi
\usepackage{url}

\usepackage{lipsum}
\graphicspath{{figures/}} %Setting the graphicspath

\newcommand{\squeezeup}{\vspace{-2.5mm}}
\newcommand{\squeezeupBY}[1]{\vspace{-#1mm}}

\newacronym{e2e}{E2E}{End-to-End}
\newacronym{lln}{LLN}{Low-power Lossy Networks}
\newacronym{m2m}{M2M}{Machine-to-Machine}
\newacronym{iot}{IoT}{Internet of Things}
\newacronym{rpl}{RPL}{Routing Protocol for Low Power and Lossy Networks}
\newacronym{ids}{IDS}{Intrusion Detection System}
\newacronym{rfid}{RFID}{Radio Frequency Identification}
\newacronym{wsn}{WSN}{Wireless Sensor Network}
\newacronym{6lowpan}{6LoWPAN}{IPv6 over Low-powered Wireless Personal Area Network}
\newacronym{ietf}{IETF}{Internet Engineering Task Force}
\newacronym{ipv6}{IPv6}{Internet Protocol version 6}
\newacronym{tcp}{TCP}{Transmission Control Protocol}
\newacronym{of}{OF}{Objective Function}
\newacronym{dag}{DAG}{Directed Acyclic Graph}
\newacronym{dodag}{DODAG}{Destination Oriented Directed Acyclic Graph}
\newacronym{ch}{CH}{Cluster Head}
\newacronym{roll}{ROLL}{Routing Over Low-power and Lossy Networks working group}
\newacronym{of0}{OF0}{Objective Function Zero}
\newacronym{mrhof}{MRHOF}{Minimum Rank with Hysteresis Objective Function}
\newacronym{ext}{EXT}{Expected Transmission Count}
\newacronym{extof}{EXTOF}{Expected Transmission Count Objective Function}
\newacronym{lbof}{LBOF}{Load Balancing Objective Function}
\newacronym{cnc}{CNC}{Child Node Count}
\newacronym{taof}{TAOF}{Traffic Aware Objective Function}
\newacronym{ptr}{PTR}{Packet Transmission Rate}
\newacronym{dodagid}{DODAGID}{DODAG Identification}
\newacronym{dio}{DIO}{DODAG Information Object}
\newacronym{dis}{DIS}{DODAG Information Solicitation}
\newacronym{dao}{DAO}{Destination Advertisement Object}
\newacronym{daoack}{DAO-ACK}{DAO Acknowledgement}
\newacronym{P2P}{P2P}{Point to Point communication}
\newacronym{p2p}{P2P}{Point-to-Point}
\newacronym{P2MP}{P2MP}{Point to Multi-Point communication}
\newacronym{p2mp}{P2MP}{Point-to-Multi-Point}
\newacronym{MP2P}{MP2P}{Multi-Point to Point communication}
\newacronym{mp2p}{MP2P}{Multi-Point-to-Point}
\newacronym{cc}{CC}{Consistency Check}
\newacronym{icmp}{ICMPv6}{Internet Control Message Protocol}
\newacronym{ccm}{CCM}{Counter wih CBC-MAC "Cipher Block Chaining - Message Authentication Code"}
\newacronym{aes}{AES}{Advanced Encryption Standard}
\newacronym{macl}{MAC}{Medium Access Control}
\newacronym{macc}{MAC}{Message Authentication Code}
\newacronym{rsa}{RSA}{Rivest-Shamir-Adleman encryption}
\newacronym{sha}{SHA}{Secure Hash Algorithm}
\newacronym{os}{OS}{Operating System}
\newacronym{gps}{GPS}{Global Positioning System}
\newacronym{rtt}{RTT}{Round-Trip Time}
\newacronym{phy}{PHY}{Physical layer}
\newacronym{coap}{CoAP}{Constrained Application Protocol}
\newacronym{ipsec}{IPSec}{Internet Protocol Security}
\newacronym{dtls}{DTLS}{Datagram Transport Layer Security protocol}
\newacronym{rssi}{RSSI}{Received Signal Stregnth Indicator}
\newacronym{rss}{RSS}{Received Signal Stregnth}
\newacronym{dos}{DoS}{Denial of Service}
\newacronym{esp}{ESP}{Encapsulated Security Header}
\newacronym{tpm}{TPM}{Trusted Platform Module}
\newacronym{ernt}{ERNT}{Extended RPL Node Trustworthiness}
\newacronym{tof}{TOF}{Trust Objective Function}
\newacronym{srpl}{SRPL}{Secure RPL}
\newacronym{dht}{DHT}{Distributed Hash Table}
\newacronym{trail}{TRAIL}{Trust Anchor Interconnection Loop}
\newacronym{vera}{VeRA}{Version attack and Rank Authentication}
\newacronym{mop}{MoP}{Mode of Operation}
\newacronym{mop2}{MoP}{mode of operation}
\newacronym{sprt}{SPRT}{Sequential Probability Ratio Test}
\newacronym{qos}{QoS}{Quality of Service}
\newacronym{ami}{AMI}{Advanced Metering Infrastructure}
\newacronym{cr}{CR}{Cognitive Radio}
\newacronym{loadng}{LOADng}{Low power and Lossy Networks On-demand Ad-hoc Distance-vector routing protocol - Next Generation}
\newacronym{loadng-ctp}{LOADng-CTP}{LOADng with Collection Tree Protocol}
\newacronym{corpl}{CORPL}{Cognitive and Opportunistic RPL}
\newacronym{aodv}{AODV}{Ad-hoc On-demand Distance Vector}
\newacronym{pdr}{PDR}{packet delivery rate}
\newacronym{pdrC}{PDR}{Packet Delivery Rate}
\newacronym{rdc}{RDC}{Radio Duty-Cycle}
\newacronym{umrpl}{UM}{Unsecured Mode}
\newacronym{psmrpl}{PSM}{Preinstalled Secure Mode}
\newacronym{asmrpl}{ASM}{Authenticated Secure Mode}
\newacronym{csmrpl}{CSM}{Chained Secure Mode}
\newacronym{sc}{SC}{Secret Chaining}
\newacronym{uc}{UC}{Unicast}
\newacronym{mc}{MC}{Multicast}
\newacronym{er}{ER}{Emergency}
\newacronym{nc}{NC}{Network Coding}
\newacronym{srr}{SRR}{SC Recovery Request/Response}
\newacronym{srreq}{SRReq}{SC Recovery Request}
\newacronym{srres}{SRRes}{SC Recovery Response}
\newacronym{bh}{BH}{Blackhole}
\newacronym{sf}{SF}{Selective-Forward}
\newacronym{sh}{SH}{Sinkhole}
\newacronym{wh}{WH}{Wormhole}
\newacronym{ra}{RA}{Rank attack}
\newacronym{va}{VA}{Version attack}
\newacronym{na}{NA}{Neighbor attack}
\newacronym{ca}{CA}{CloneID attack}
\newacronym{dtm}{DTM}{Dynamic Threshold Mechanism}
\newacronym{iana}{IANA}{Internet Assigned Numbers Authority}

\makeglossaries
\glsaddall
\IEEEaftertitletext{\vspace{-1\baselineskip}} % reducing space after authors section in journal mode

\begin{document}
\title{Securing RPL using Network Coding: The Chained Secure Mode (CSM)}

\author{Ahmed~Raoof,~\IEEEmembership{Student~Member,~IEEE,}
	Chung-Horng~Lung,~\IEEEmembership{Senior~Member,~IEEE,}
	and~Ashraf~Matrawy,~\IEEEmembership{Senior~Member,~IEEE}% <-this % stops a space
	\thanks{A. Raoof and C. Lung are with the Department of Systems and Computer Engineering,
		Faculty of Engineering and Design, Carleton University, Ottawa, ON, Canada (email: ahmed.raoof@carleton.ca; chlung@sce.carleton.ca)}% <-this % stops a space
	\thanks{A. Matrawy is with the School of Information Technology, Carleton University, Ottawa, ON, Canada (email: amatrawy@sce.carleton.ca)}% <-this % stops a space
	\thanks{Copyright (c) 2021 IEEE. Personal use of this material is permitted. However, permission to use this material for any other purposes must be obtained from the IEEE by sending a request to pubs-permissions@ieee.org.}% <-this % stops a space
}

\maketitle
%\IEEEpeerreviewmaketitle

\markboth{IEEE~Internet of Things,~Vol.~xx, No.~x, XXX~2021}{~Raoof \MakeLowercase{\textit{et al.}}: Securing RPL using Network Coding: The Chained Secure Mode (CSM)}

\begin{abstract}
Considered the preferred routing protocol for many Internet of Things (IoT) networks, the \gls{rpl} incorporates three security modes to protect the integrity and confidentiality of the routing process: the \gls{umrpl}, \gls{psmrpl}, and the \gls{asmrpl}. Both \gls{psmrpl} and \gls{asmrpl} were originally designed to protect against external routing attacks, in addition to some replay attacks (through an optional replay protection mechanism). However, recent research showed that \gls{rpl}, even when it operates in \gls{psmrpl}, is still vulnerable to many routing attacks, both internal and external. In this paper, a novel secure mode for \gls{rpl}, the \gls{csmrpl}, is proposed using the concept of intra-flow \gls{nc}. The \gls{csmrpl} is designed to enhance \gls{rpl}'s resiliency and mitigation capability against replay attacks. In addition, \gls{csmrpl} allows the integration with external security measures such as \glspl{ids}. An evaluation of the proposed \gls{csmrpl}, from a security and performance point of view, was conducted and compared against \gls{rpl} in \gls{umrpl} and \gls{psmrpl} (with and without the optional replay protection) under several routing attacks: the \gls{na}, \gls{wh}, and \gls{ca}, using average \gls{pdr}, \gls{e2e} latency, and power consumption as metrics. It showed that \gls{csmrpl} has better performance and more enhanced security than both the \gls{umrpl} and \gls{psmrpl} with the replay protection while mitigating both the \gls{na} and \gls{wh} attacks and significantly reducing the effect of the \gls{ca} in the investigated scenarios.
\end{abstract}


\section{Introduction}
The \acrfull{rpl}\cite{RFC6550} has attracted a great deal of attention since it became a standard in 2012. The security aspect of \gls{rpl} has been of a special interest, including different routing attacks the protocol is susceptible to\cite{Raoof2018, Perazzo2018, Aris2016}, mitigation methods and \acrfullpl{ids}\cite{Dvir2011, Gara2017, Perrey2016}, and performance evaluation of some of \gls{rpl}'s security mechanisms\cite{Raoof2020, Raoof2019a, Arena2020}.

Our previous work\cite{Raoof2019a, Raoof2020} showed that \gls{rpl}'s secure modes, while providing reasonable mitigation of some external attacks, are still susceptible to many routing attacks (both internal and external) (see \S\ref{CSMmotiv}), especially the replay attacks such as \acrfull{na} and \acrfull{wh} attack.

This paper presents a significant extension to our previous conference paper\cite{Raoof2020a}, where we devised a proof-of-concept prototype of \acrfull{csmrpl}. In this work, We added a proper \gls{sc} recovery mechanism to \gls{csmrpl} and the capability to integrate external security measures (e.g., \glspl{ids}) into \gls{csmrpl}. Then, a thorough evaluation of the improved \gls{csmrpl} was executed against the other \gls{rpl} secure modes in the presence of several routing attacks - see \S\ref{CSMEval}.

Our contributions can be summarized as follows:
\begin{itemize}
	\item A novel secure mode for \gls{rpl}, the \gls{csmrpl}, was designed and complemented with a proper recovery mechanism and integration capability with external security mechanisms. \gls{csmrpl} makes use of the idea of intra-flow \gls{nc} to create a linked chain of coded \gls{rpl} control messages between every two neighboring nodes (see \S\ref{NC}). The effect of the linked chain can limit adversaries' ability to launch some routing attacks, e.g., identity-cloning and replay attacks (e.g., \gls{wh} and \gls{na} attacks)\cite{Raoof2018}.
	\item A prototype of the proposed \gls{csmrpl} was designed and implemented in Contiki \gls{os}\cite{ContikiOSRef}, including the newly-added features mentioned above.
	\item Using 350 simulation experiments, and to demonstrate the capabilities of the \gls{csmrpl} prototype, a security and performance comparison between \gls{rpl} in \gls{csmrpl} and \acrfull{psmrpl} (with and without the optional replay protection) against the \gls{na}, \acrfull{ca}, and \gls{wh} attacks was conducted using several metrics.
	\item For the internal adversary cases, the results showed that \gls{csmrpl} is capable of mitigating both the \gls{na} and \gls{wh} attacks with less latency ($\approx$95\% less) and power consumption ($\approx$13-28\% less) than \gls{psmrpl} with replay protection. In addition, \gls{csmrpl} showed enhanced security and was able to significantly reduced the impact of \gls{ca} on the network (\gls{pdr} is 5-18\% higher, \gls{e2e} latency is $\approx$95\% less, and with comparable power consumption to that of \gls{psmrpl}, compared to all other secure modes. See \S\ref{INTResults}.)
	\item For the external adversary cases, the results showed that \gls{csmrpl} is the only secure mode capable of mitigating the \gls{wh} attack with \gls{pdr} $\approx$95-99\%, \gls{e2e} latency between 5-10 milliseconds, and power consumption similar to the normal operation of \gls{rpl} in \gls{umrpl}. See \S\ref{EXTResults}.
\end{itemize}

The remainder of this paper is structured as follows: the next section describes the related works. Section \ref{background} presents an overview of \gls{rpl} and its security mechanisms. Section \ref{CSM} presents the proposed \gls{csmrpl}. The experimental setup and assumptions are described in section \ref{CSMEval}. Sections \ref{INTResults} and \ref{EXTResults} present the evaluation results and analysis, respectively, which is followed by a discussion in Section \ref{Obsr}. Section \ref{Conc} presents the conclusion drawn from the proposed \gls{csmrpl}.
\squeezeup
\section{Related Works}\label{rltdwrk}
An implementation of \gls{psmrpl} for \gls{rpl} was provided by Perazzo \textit{et al.} in \cite{Perazzo2017}, along with the optional replay protection, the \gls{cc} mechanism. This implementation was based on ContikiRPL (Contiki OS version of \gls{rpl}). The authors evaluated their implementation, and compared \gls{rpl}'s performance between \gls{psmrpl} and \gls{umrpl}. Their evaluation results showed that the replay protection mechanism introduced higher network formation time and increased power consumption. In an effort to enhance the \gls{cc} mechanism, an optimized version of it was introduced in \cite {Arena2020} that uses \gls{rpl} options\cite{RFC6550} to include another unique nonce value within the exchanged \gls{cc} messages. The evaluation of the optimized mechanism showed a 36\% shorter network formation time and 45\% decrease in the \gls{cc} messages exchanged while maintaining the same level of protection. It was shown in our work in \cite{Raoof2020} that, based on the authors' implementation, \gls{psmrpl}rp is still vulnerable to replay attacks, specifically the \gls{wh} attack where the adversaries will replay the \gls{cc} messages between the victim nodes.

Airehrour \textit{et al.} in \cite{Airehrour2019} proposed a modified version of \gls{rpl}, named \textit{SecTrust-RPL}, which used their devised SecTrust framework\cite{Airehrour2018}. In the SecTrust framework, the optimum route is chosen based on the trust evaluation of the nodes, resulting in isolating suspected adversaries. Trust is calculated based on the successful packet exchange between the nodes, and it is dependent on time. \textit{SecTrust-RPL} was evaluated under the Decreased Rank and Sybil attacks using Contiki \gls{os} in both simulation and a real testbed. Compared to \gls{rpl} in \gls{umrpl} under the same attacks, \textit{SecTrust-RPL} showed a significant decrease in lost packets ($\approx$60\%) and lower rank changes among the nodes. However, the authors did not evaluate the effect of their implementation on power consumption and the \gls{e2e} latency.

\section{RPL Brief Overview}\label{background}
Being a distance-vector routing protocol, \gls{rpl} arranges the network nodes into a \gls{dodag}\cite{Janicijevic2011}: a network of nodes connected without loops with the traffic directed toward one \textit{root} node\cite{RFC6550,Granjal}. The creation of \glspl{dodag} and how parents are selected depends on two aspects: the \textit{\gls{of}}, which defines the used \gls{rpl} configurations, and the node's \textit{rank}\footnote{The rank of a node represents its distance to the root node based on the routing metrics defined by the \gls{of}}.

The \gls{dodag} in \gls{rpl} is built by exchanging control messages, which have five types; four of them have two versions (base and secure versions), and the last one has only a secure version. Enabling any of the secure modes of \gls{rpl} (explained later in this section) switches the control messages to their secure versions. The secure control messages add new unencrypted header fields and either a \gls{macc} or a digital signature field to the end of the base version, then encrypts the base part and the \gls{macc}/signature field\cite{RFC6550}. As stated in the \gls{rpl} standard\cite{RFC6550}, \gls{rpl} messages are sent as \gls{icmp} messages, with the "\textit{Type}" field in its header equal to 155 -- as set by \gls{iana} -- and the "\textit{Code}" field identifying the type of the \gls{rpl} control message\cite{RFC6550}.

The five types of \gls{rpl} control messages go as follows\cite{RFC6550}: \gls{dio} and \gls{dis} messages are used for the creation and maintenance of the upward \gls{dodag}, while the \gls{dao} / \gls{daoack} pair of messages are used to create the downward routing table. Finally, the \gls{cc} messages are the basis for the optional replay protection mechanism, where non-repetitive nonce values are exchanged and used to assure no \gls{dio} message replay had occurred\cite{RFC6550, Airehrour2019}.

\gls{rpl} standard currently offers three security modes to ensure control messages' confidentiality and integrity\cite{RFC6550, Perazzo2017}: (i) \textit{\textbf{\gls{umrpl}}}, where no \gls{rpl} security features are enabled and only the link-layer security is applied, if available (default mode); (ii) \textbf{\textit{\gls{psmrpl}}}, in which preinstalled symmetrical encryption keys are used to encrypt \gls{rpl} control messages; and (iii) \textit{\textbf{\gls{asmrpl}}} where two keys are used within the network: preinstalled keys are used by nodes for joining the network as leaf nodes, after that all routing-capable nodes must acquire new keys from an authentication authority after being authenticated. The new keys are used between the routing nodes only. However, the \gls{rpl} standard\cite{RFC6550} leaves all the details of \gls{asmrpl} (e.g., how the authentication is performed, the exchange of the new encryption keys, etc.) for a future specification that has not been worked out yet.

As an optional security mechanism that is only available in the preinstalled (PSMrp) or authenticated mode (ASMrp), \gls{rpl} offers a replay protection mechanism called the \acrlong{cc}. In these checks, special secure control messages (\gls{cc} messages) with non-repetitive nonce value are exchanged and used to assure no message replay had occurred\cite{RFC6550, Airehrour2019}.

It is worth mentioning that all of the popular \gls{iot} operating systems (e.g., Contiki \gls{os}\cite{ContikiOSRef} and TinyOS\cite{TinyOS}) have implemented \gls{rpl} in \gls{umrpl} only. To the best of our knowledge, \gls{asmrpl} has never been implemented, and it was not until recently that \gls{psmrpl} was implemented by Perazzo \textit{et al.}\cite{Perazzo2017}, albeit in an experimental form.

\section{The Proposed Chained Secure Mode (CSM)}\label{CSM}
\subsection{Motivations}\label{CSMmotiv}
Our work in \cite{Raoof2019a, Raoof2020} examined \gls{rpl} secure modes' performance under several routing attacks, and have shown that \gls{psmrpl} (and by extension, \gls{asmrpl}) can mitigate most of the external attacks\footnote{External attack is launched by an adversary who is not part of the network, e.g., it does not have the encryption key used by the legitimate nodes for \gls{rpl} in \gls{psmrpl}, or runs \gls{rpl} in \gls{umrpl}.}. However, it does not increase \gls{rpl}'s security against internal attacks\footnote{An internal attack is launched by an adversary who is part of the network, e.g., it has the encryption key used by the legitimate nodes for \gls{rpl} in \gls{psmrpl}.}. In addition, we showed that replay attacks could still be triggered by external adversaries, even when \gls{psmrpl}rp is used (e.g., in the case of the \gls{wh} attack.)

Further, we have investigated \gls{rpl} standard\cite{RFC6550} and found out that it only provides confidentiality and integrity of its control messages, without any verification of their sender's authenticity. It means that the door is wide open for attacks such as the Sybil, identity-cloning, eavesdropping, and replay attacks\cite{Raoof2018} to be launched regardless of the secure mode \gls{rpl} is running. For example, a \textbf{\textit{\acrlong{na}}} - see \S\ref{AttMdl} - can be easily launched from an external adversary simply by checking the "\textit{Type}" and "\textit{Code}" fields in any \gls{icmp} message header to identify \gls{rpl}'s \gls{dio} messages\footnote{(Type = 155) means this is an \gls{rpl} message. (Code = 1 or 129) means it is a regular or secure \gls{dio} message, respectively.}, without the need to decrypt the actual message\cite{Raoof2020}.
\begin{figure*}[ht]
	\centering
	\subfloat[Normal network communication]{\includegraphics[scale=0.5]{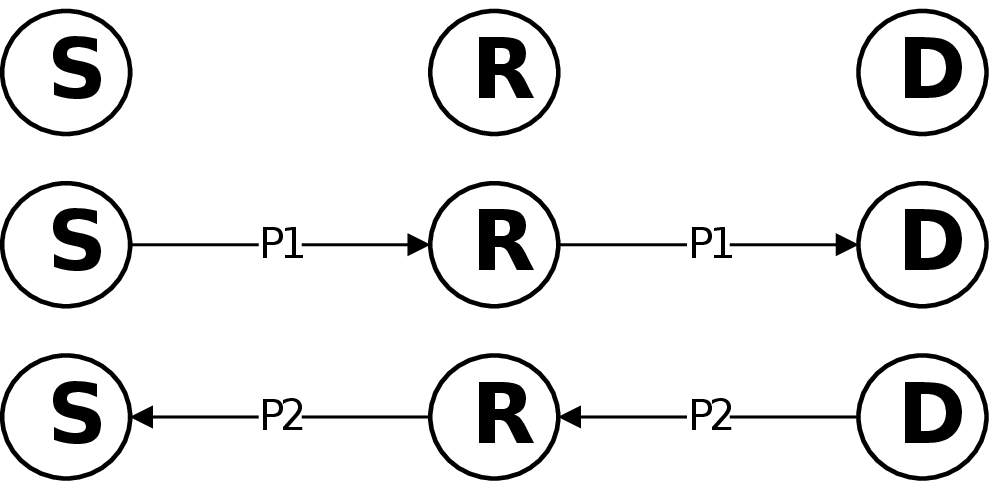}%
		\label{fig_NC_NoNC}}
	\hfil
	\subfloat[Inter-flow \gls{nc}]{\includegraphics[scale=0.5]{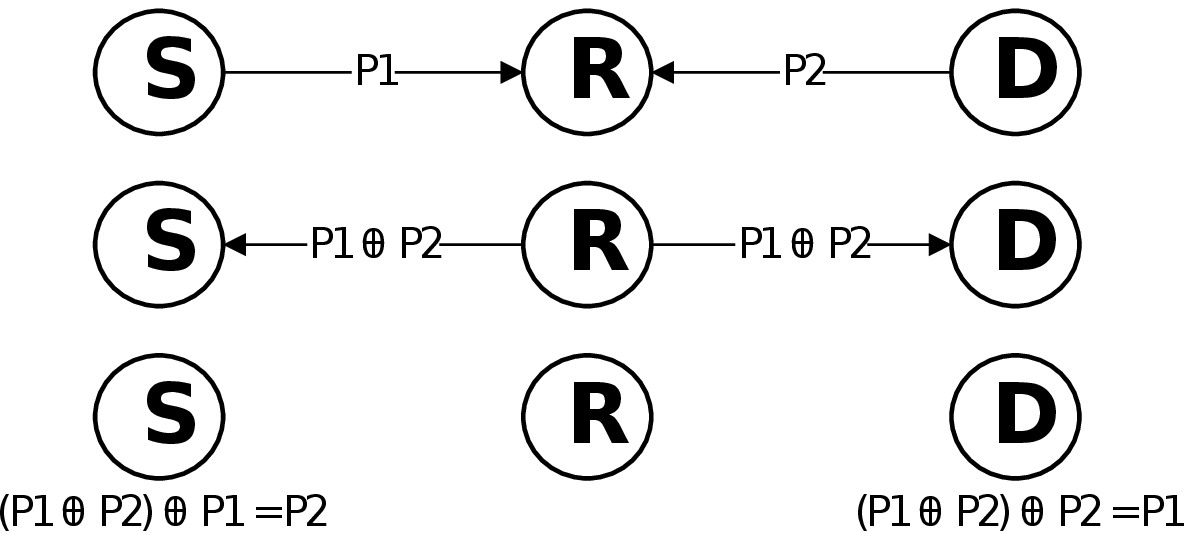}
		\label{fig_NC_InterFlowNC}}
	\hfil
	\subfloat[Intra-flow \gls{nc}]{\includegraphics[scale=0.5]{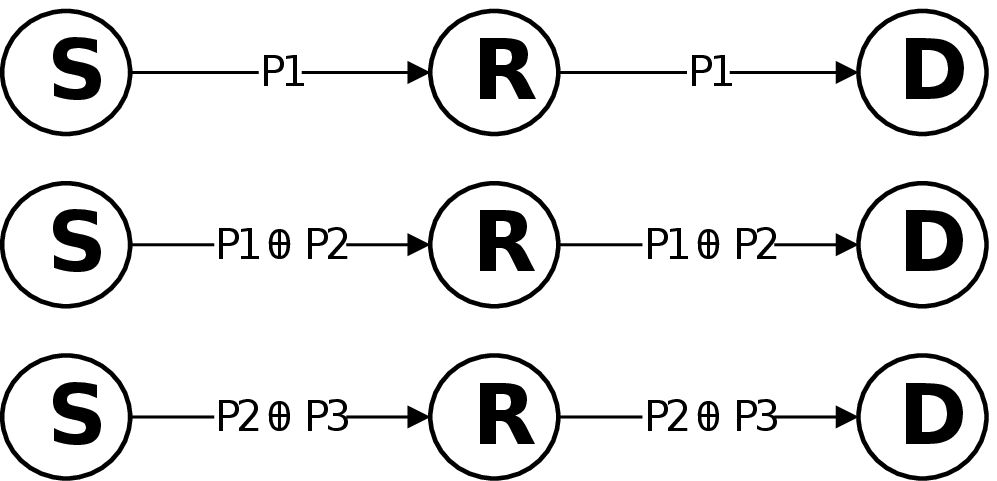}
		\label{fig_NC_IntraFlowNC}}
	\caption{Examples of \gls{nc} communication. The $ \oplus $ sign represents XOR as a simple \gls{nc} operation.}
	\label{fig_NC}
	\squeezeup
\end{figure*}

The lack of sender authentication in \gls{rpl} control messages motivated us to devise an innovative method to overcome this problem. Integrating the intra-flow \gls{nc} scheme into \gls{rpl} provides a proof of message authenticity for any receiving node, assuming that the first message truly came from the original sender. This scheme stands true for most attacks as the adversaries mostly join the network after it has been initiated and established.

\begin{table}[!t]
	\caption{List of Used Abbreviations for \gls{csmrpl}}
	\label{table_ABB}
	\begin{tabular}{ll}
		\toprule
		\textbf{Abbreviation}&\textbf{Description}\\
		SC&Secret Chaining\\
		xC&Either Unicast (\textbf{UC}) or Multicast (\textbf{MC}) flow\\
		ER&Emergency flow\\
		TX&Transmitting\\
		RX&Receiving\\
		SRRxx&
		\begin{minipage}{0.72\columnwidth}
			SC Recovery control message, either Request (SRReq) or Response (SRRes)
		\end{minipage} \\
		\bottomrule
	\end{tabular}
\end{table}
\squeezeup
\subsection{Brief Review on Network Coding}\label{NC}
Since its first proposal be Ahlswede \textit{et al.}\cite{Ahlswede2000}, \gls{nc} has received a great deal of research attention, as many researchers investigated \gls{nc} schemes (e.g., \textit{XOR}, \textit{Random Linear \gls{nc}}, etc.) to improve network efficiency (e.g., throughput, reliability, and \gls{e2e} delay) on different communication technologies (wired, wireless, or ad hoc networks)\cite{Hay2014}.

\gls{nc}'s basic idea is that a source combines multiple pieces of information or packets using a coding scheme and forwards the coded information to the next network device. At the receiver's end, and upon receiving enough information, the combined information is decoded to recover the original data. 

The simplest \gls{nc} scheme is XOR. For example, a device can perform bit-by-bit XOR operations of two packets in sequence and forward the XOR-ed packet to the next hop to reduce the number of transmissions\cite{Ahlswede2000}.

Implementation-wise, \gls{nc} can be applied to either (i) \textit{inter-flow} traffic; for which \gls{nc} applies coding to packets from different traffic flows (see Fig. \ref{fig_NC_InterFlowNC}), or (ii) \textit{intra-flow} traffic, whereas \gls{nc} applies coding to packets of the same traffic flow\cite{Hansen2018, Saeed2013} (see Fig. \ref{fig_NC_IntraFlowNC}), creating a \textit{chain} of messages. Inter-flow \gls{nc} requires more complex operations, such as buffering and synchronization of packets from multiple flows or different sources. Intra-flow \gls{nc}, on the other hand, is much easier as it only considers the sequence of packets within the same flow, which makes it suitable to the resource-constrained \gls{iot}.

This paper proposes an innovative secure mode for \gls{rpl}, the \gls{csmrpl}, using the intra-flow \gls{nc}. The \textit{chaining} effect from this method adds sender authenticity to \gls{rpl} (assuming that the first message came from the original sender) and increases its resilience against several routing attacks (e.g. replay attacks). For concept demonstration only, we make use of the simplest \gls{nc} scheme, the \textit{\textbf{XOR}}. However, more sophisticated \gls{nc} schemes can be used for higher level of security.

Throughout this paper, a \textit{flow} is defined as the stream of \gls{rpl} control messages from a node toward a specific \gls{ipv6} address. This can be for a \gls{uc} transmission (a unicast \gls{ipv6} address of a certain neighbor of the node) or a \gls{mc} transmission (a multicast \gls{ipv6} address.)

\subsection{How CSM Operates}\label{CSMwork}
First, the used abbreviations are listed in Table \ref{table_ABB}. The design of \gls{csmrpl} is based on the following points:
\begin{itemize}
	\item Adhering to the \gls{rpl} standard by maintaining the same procedures used for \gls{psmrpl}.
	\item Using intra-flow \gls{nc} to provide broad replay-attacks-mitigation capability as part of \gls{rpl} standard.
	\item Allowing external security measures to integrate with \gls{csmrpl} by controlling how \gls{rpl} trusts nodes.
	\item \gls{csmrpl}'s primary focus is on protecting static networks, as they constitute the majority of current \gls{iot} applications. However, mobility can be supported using the external security measure integration mentioned above.
\end{itemize}

\begin{figure*}[!t]
	\centering
	\includegraphics[scale=0.65]{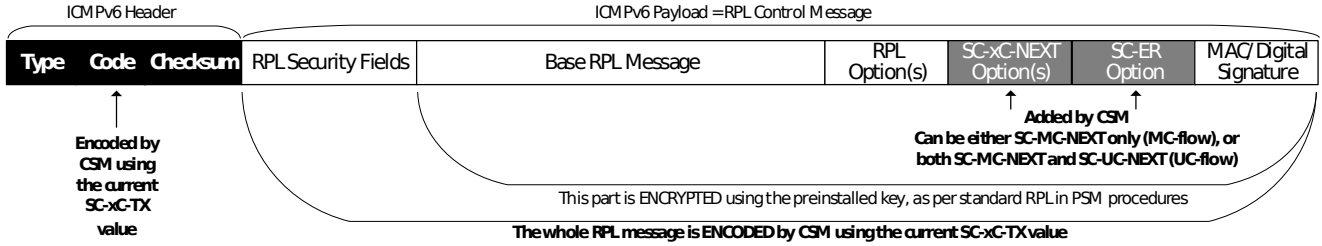}%
	\caption{Format of an \gls{rpl} control message, as constructed by the proposed \gls{csmrpl}. The black parts represents \gls{icmp} header, the white parts are standard \gls{rpl} in \gls{psmrpl} fields, and the grey parts are added by \gls{csmrpl}.}
	\label{fig_CSM_RPLMsg}
	\squeezeup
\end{figure*}
\begin{figure}[!t]
	\centering
	\subfloat[Sending an \gls{rpl} message]{\includegraphics[scale=0.49,valign=c]{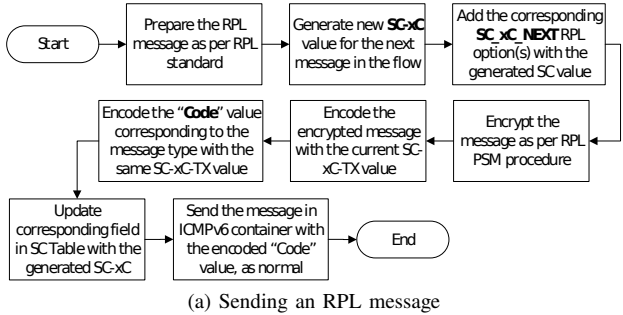}%
		\label{fig_FC_SndProc}}
	\hfil
	\subfloat[Receiving an \gls{rpl} message]{\includegraphics[scale=0.55,valign=c]{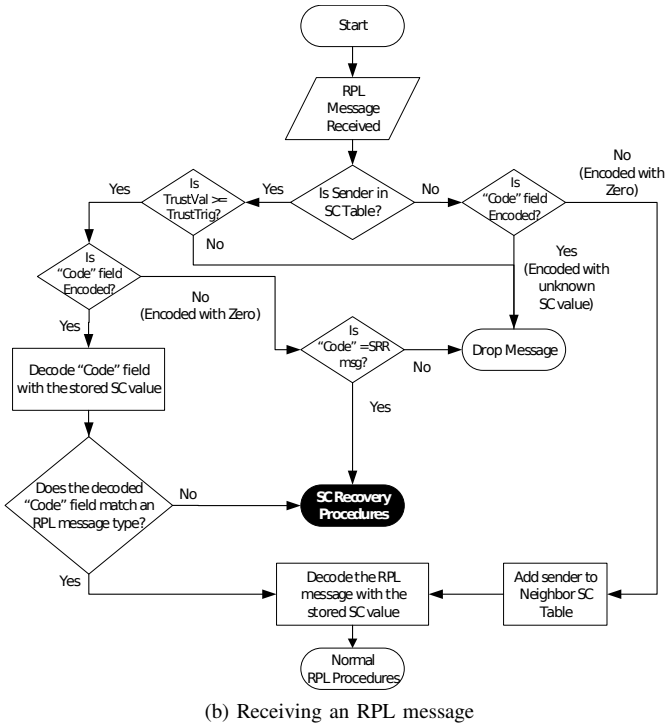}
		\label{fig_FC_RcvProc}}
	\caption{Flowcharts represent the sending and reception procedure of an \gls{rpl} message in the current \gls{csmrpl} prototype.}
	\label{fig_FC_CSM}
	\squeezeup
\end{figure}
\begin{figure*}[!t]
	\centering
	\includegraphics[scale=0.59]{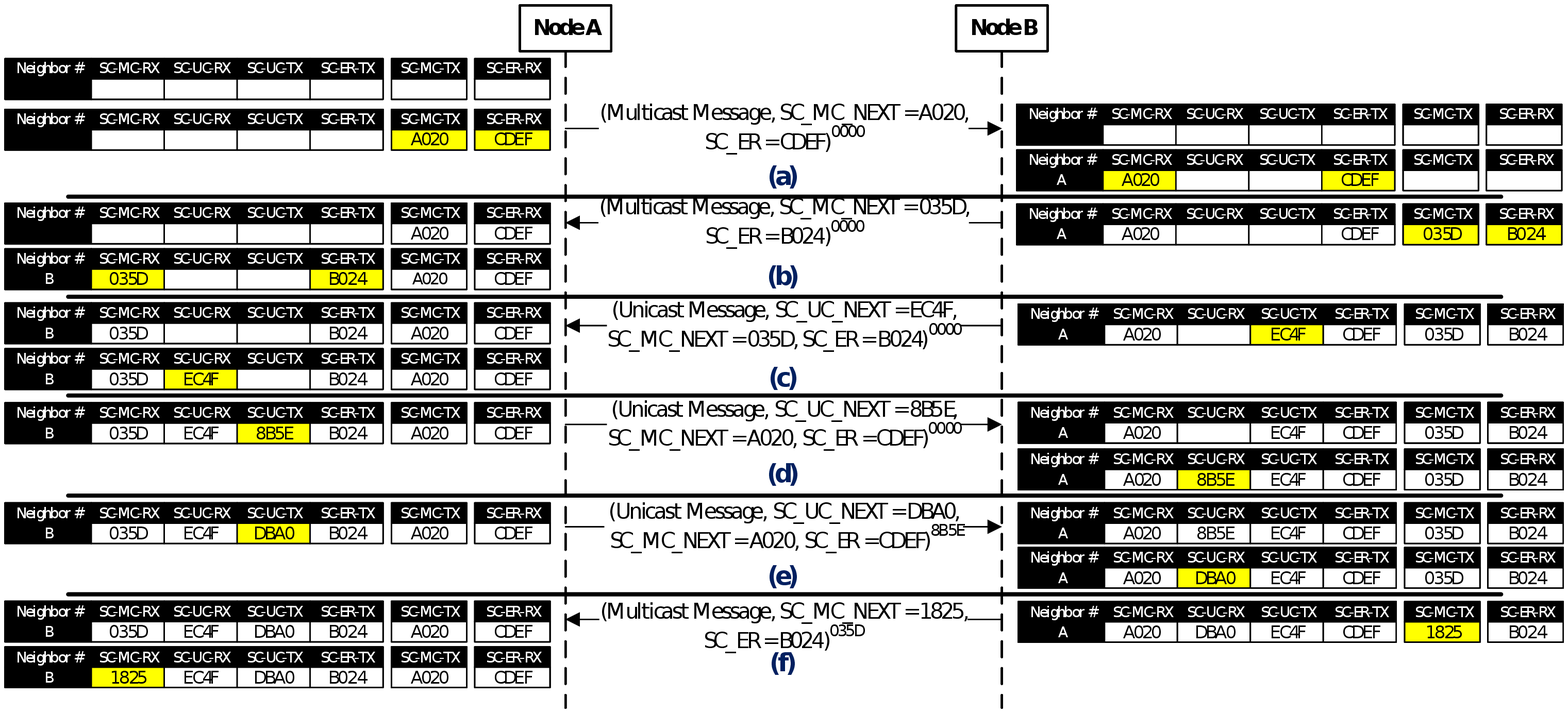}%
	\caption{Examples of normal \gls{csmrpl} operation in chronological order (the number on the top-right of the brackets represents the \gls{sc} value used to encode that message): (a and b) the first message in the \gls{mc}-flow, (c and d) the first message in the \gls{uc}-flow, (e) subsequent messages of the \gls{uc}-flow, and (f) subsequent messages of the \gls{mc}-flow. The yellow color highlights a creation or a change of an \gls{sc} value in the \gls{sc} table.}
	\label{fig_CSM}
	\squeezeup
\end{figure*}

To implement the intra-flow \gls{nc}, and instead of using the entire previous control message for the \textit{encoding} / \textit{decoding} of the current one (see Fig. \ref{fig_NC_IntraFlowNC}), \gls{csmrpl} uses the \textbf{\acrfull{sc}} values which are sent within the previous control message. These \gls{sc} values are 4-byte unsigned, randomly generated integer numbers (for each sent control message), and are locally unique for each neighbor.

Since \gls{rpl} sends its control messages as either \acrfull{mc} or \acrfull{uc} messages, \gls{csmrpl} considers them two independent flows: an \gls{mc}-flow and a \gls{uc}-flow. Hence, every node in the network should maintain a table (the \textit{\gls{sc} table}) of the following \gls{sc} values for each neighbor, in order to successfully encode and decode their control messages:
\begin{itemize}
	\item \textbf{SC\_UC\_RX:} The \gls{sc} value used to decode the next incoming \gls{uc}-flow message from the neighbor.
	\item \textbf{SC\_MC\_RX:} The \gls{sc} value used to decode the next incoming \gls{mc}-flow message from the neighbor.
	\item \textbf{SC\_UC\_TX:} The \gls{sc} value used to encode the next outgoing \gls{uc}-flow message to the neighbor.
	\item \textbf{SC\_ER\_TX:} The \gls{sc} value used to encode the next outgoing \gls{er}-flow message to the neighbor - see \S\ref{SCRcvry}.
\end{itemize}

In addition, each node should maintain the next \gls{sc} value for its next \gls{mc}-flow transmission (\textbf{SC\_MC\_TX}) and \gls{er}-flow reception (\textbf{SC\_ER\_RX}) -- see \S\ref{SCRcvry}. For simplicity, the current \gls{csmrpl} design uses \textit{zero} as a value for the \gls{sc} used for the first transmission in each flow.

To exchange the \gls{sc} values used to encode the next control message, \gls{csmrpl} employs the \textit{\gls{rpl} Control Message Options} from the standard\cite{RFC6550}. These optional add-ons are used to provide (or request) information to (or from) the receiver. \gls{csmrpl} adds three new options to accommodate the transmission of the next \gls{sc} used for each flow: the (\textbf{SC\_UC\_NEXT}) option includes the \gls{sc} value to be used for the next \gls{uc}-flow message, (\textbf{SC\_MC\_NEXT}) is for the \gls{sc} value to be used for the next \gls{mc}-flow message, and (\textbf{SC\_ER}) is for the \gls{sc} value to be used for the next \gls{er}-flow message -- see \S\ref{SCRcvry}.

When a node wants to send an \gls{rpl} control message (whether for the \gls{uc}-, \gls{mc}, or \gls{er}-flow), it will follow the following steps (see Fig. \ref{fig_FC_SndProc}):
\begin{enumerate}
	\item Prepare the message as per the standard \gls{psmrpl} procedures.
	\item Generate a new \gls{sc} value to be sent within the corresponding \gls{rpl} option. The generation process also ensures that the generated \gls{sc} values, when used to encode the \gls{icmp} Code field (see step 4), will not result in one of the valid values of the \gls{rpl} message type identifier. This step is not performed for \gls{srr} messages - see \S\ref{SCRcvry}.
	\item Adding the (SC\_xC\_NEXT) and (SC\_ER) new control message options, as per the \gls{rpl} standard. \gls{csmrpl} should add both the (SC\_UC\_NEXT) and (SC\_MC\_NEXT) for \gls{uc}-flow messages and only the (SC\_MC\_NEXT) for the \gls{mc}-flow messages. The use of both options for the \gls{uc}-flow allows for quicker recovery from message chain breakage in the \gls{mc}-flow.
	\item The \textit{Code} field of the \gls{icmp} header is encoded using the corresponding SC\_UC\_TX or SC\_MC\_TX value to mitigate the security vulnerability addressed in \S\ref{CSMmotiv}. The only exception is the \textit{\gls{srr}} messages, which keep their designated \gls{icmp} "Code" value without encoding -- see \S\ref{SCRcvry}. Since  the Code field is one byte long, its encoding will be a sequential one, i.e., it is encoded first with the first byte of the \gls{sc} value, then the result is encoded with the second byte of the \gls{sc} value, and so on.
\end{enumerate}

After encrypting the message (according to standard \gls{psmrpl} procedures), \gls{csmrpl} will encode the whole message using the corresponding \gls{sc} value then send it as usual. Fig. \ref{fig_CSM_RPLMsg} depicts how \gls{csmrpl} constructs an \gls{rpl} message, while Fig. \ref{fig_FC_SndProc} represents a flowchart of \gls{rpl} message sending procedure in \gls{csmrpl}.

At the receiving node, the decoding \gls{sc} value is found from the \gls{sc} table using the sender IP address. The found \gls{sc} value is used to decode the \textit{Code} field of the \gls{icmp} header to identify the type of \gls{rpl} message. The whole message then is decoded using the same \gls{sc} value and is processed based on \gls{psmrpl}'s procedures. If the \textit{Code} field cannot be decoded, the message will be discarded without processing. Fig. \ref{fig_FC_RcvProc} shows a flowchart for message reception in \gls{csmrpl}.% Fig. \ref{fig_CSM} shows examples of \gls{csmrpl} normal operation.

Fig. \ref{fig_CSM} shows a few examples of the normal \gls{csmrpl} operations. Parts (a) and (b) represent the first \gls{mc} transmissions, where the sender node generates a new SC\_MC\_TX value (to be used for its next \gls{mc} transmission) and SC\_ER\_RX. Then, it adds the newly-introduced SC\_xC\_NEXT and SC\_ER \gls{rpl} control message options (which includes the generated \gls{sc} values) to the sent message before encoding it with zero (this is the very first \gls{mc} transmission). At the receiver side, once the message is decoded successfully (using zeros for the \gls{sc} value as it is the first \gls{mc} message from the sender), the receiver adds an entry to its SC Table for the sender with the \gls{sc} values it extracts from the received message, e.g., node B will store the value A020 in the SC\_MC\_RX field of node A's entry. A similar situation is shown in parts (c) and (d) for the first \gls{uc} transmission. However, the difference here is the inclusion of both the \gls{uc} and \gls{mc} \gls{sc} values in the sent message, which means all \gls{sc} values of the sender will be updated. Finally, parts (e) and (f) show how the subsequent \gls{uc} and \gls{mc} transmissions will update the SC Table at the nodes. For example, in part (e), node B will use the stored SC\_UC\_RX for node A (8B5E) to decode the received \gls{uc} message, then updates node A's entry with the extracted \gls{sc} values, e.g., SC\_UC\_RX will become DBA0.

\subsection{SC Recovery Mechanism}\label{SCRcvry}
Our initial work\cite{Raoof2020a} showed that \gls{csmrpl} requires a proper recovery mechanism when a control message from any \gls{nc} flow is missed or lost, otherwise all subsequent communications in that flow will be discarded due to not having the correct \gls{sc} value to decode it.

As \gls{csmrpl} is designed for the static networks, there are two cases where nodes can loose track of the \gls{sc} values:-
\begin{enumerate}
	\item \textbf{Node Reset:} When a node is reset for any reason (e.g., battery replacement, firmware upgrades, etc.), it will lose all stored \gls{sc} values and start from scratch. For this, \gls{csmrpl} assumes that the \gls{os} will periodically save all \gls{sc} values (node's own and \gls{sc} table) to the filesystem, and loads them at boot-up time, so the node can resume from the point just before the reset.
	\item \textbf{Lost or Corrupt Messages:} Missing a control message is normal for lossy networks such as the ones \gls{rpl} is designed for\cite{RFC6550}. However, in \gls{csmrpl} this means breaking the message-chain for one of (or both) flows, resulting in discarding all subsequent messages of the broken flow.
\end{enumerate}

To recover from the second case mentioned above, \gls{csmrpl} implements a special recovery mechanism, dubbed the \textit{\gls{sc} Recovery}. This recovery mechanism applies only to the neighbors that are already in the \gls{sc} table. To secure the recovery process, the \gls{sc} values are not sent as clear text. Instead, a challenge/response exchange is performed based on the concept of solving Linear Algebra equations that involves the missing \gls{sc} values. In general, assuming that node (A) is the receiver of the "non-decodable" message and node (B) is the original sender, node (A) will send an \gls{srreq} message to node (B) containing the coefficients of a system of linear equations, to which node (B) will use the coefficients and its next \gls{sc} values to calculate the results of the linear equations. These results are replied to node (A) inside an \gls{srres} message. Now, node (A) will use the provided information to solve the linear equations and extracts the missing \gls{sc} values for node (B). The reason behind this procedure is to raise the bar for the adversaries to launch attacks against (or abusing) the recovery mechanism.

Based on the above-mentioned concept, The \gls{sc} recovery mechanism goes as follows:
\begin{itemize}
	\item A new \gls{nc} flow is added to \gls{csmrpl}, the \textit{Emergency (ER) flow}, to be used when exchanging the \gls{srr} messages. Each node will maintain an \gls{sc} value (\textbf{SC\_ER}) for this flow, and exchanges it through the (\textbf{SC\_ER}) option in every message of the other flows -- see \S\ref{CSMwork}. However, the SC\_ER only updates after a successful recovery.%\gls{sc} recovery attempt.
	\item When a message is received, if the decoded "Code" field at the \gls{icmp} header of the received message does not represent an \gls{rpl} control message, the following methods are performed to recover the missing \gls{sc} value:
		\begin{itemize}
			\item If the received message was from the \gls{mc}-flow, a regular \gls{uc}-\gls{dis} message is sent to the sender. As per \gls{rpl} standard, the sender must reply with a regular \gls{uc}-\gls{dio}, which will have all the correct \gls{sc} values for the next message in all the flows.
			\item If the received message was from the \gls{uc}-flow, the \gls{sc} recovery mechanism is conducted:
			\begin{enumerate}
				\item The received message will be discarded.
				\item The receiver sends a \gls{uc}-\gls{srreq} message to the original sender, containing the randomly-generated coefficient values to be used by the original sender as explained above. The \gls{srreq} is encoded with the original sender's \textbf{SC\_ER} value.
				\item Once receiving the \gls{srreq} message, the original sender will calculate the results of the linear equations, then it sends them within a \gls{mc}-\gls{srres} message to the receiver, encoded with the receiver's \textbf{SC\_ER} value. Since this is a multicast message, the \gls{srres} message contains the \gls{ipv6} address of the node that is supposed to process it. In addition, the original sender will update his SC\_ER and send it within the corresponding \gls{rpl} option.
				\item Now, the receiver will use the coefficients and the received results to solve the linear equations and update its \gls{sc} table with the extracted \gls{sc} values.
			\end{enumerate}
		\end{itemize}
\end{itemize}

\subsection{The CSM-Trust Integration Interface}\label{CSM-TrustValIntfc}
To allow integration with external security measures, \gls{csmrpl} provides a trust-based control interface for such security systems called the \textit{\gls{csmrpl}-Trust} interface, which uses the \textit{TrustVal} value (as part of the \gls{sc} table) to define the trust-worthiness of each of the node's neighbors. Hence, the acceptance or rejection of \gls{rpl}'s control messages from that neighbor.
\begin{figure}[!t]
	\centering
	\includegraphics[scale=0.40]{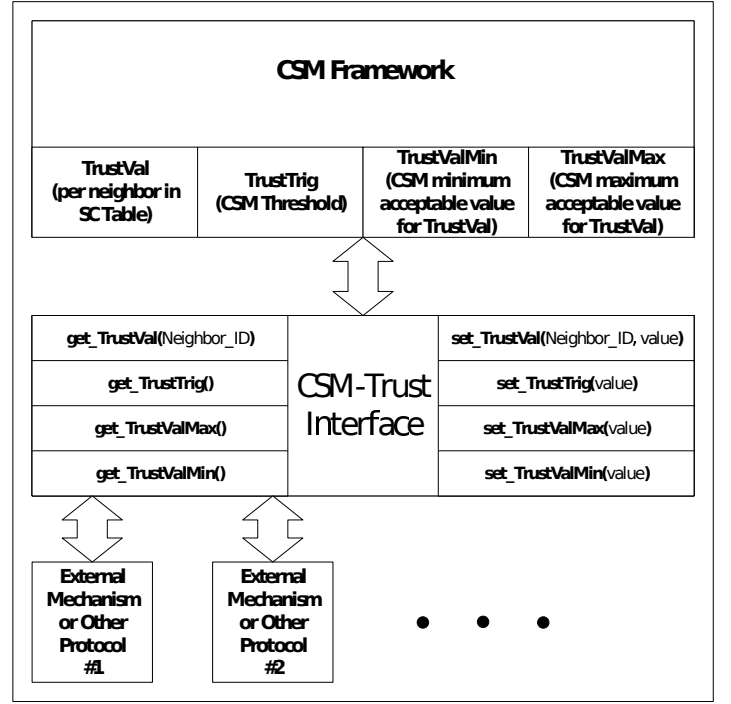}%Topology-All}%[height=65mm, width=70mm]
	\caption[\textit{CSM-Trust} interface conceptual diagram.]{\textit{CSM-Trust} interface conceptual diagram.}
	\label{fig_CSM-Trust}
\end{figure}

The \textit{\gls{csmrpl}-Trust} interface allows external security measures to set and use the \textit{TrustVal} value according to their needs. For example, an external security mechanism (such as an \gls{ids}) would set the boundaries for \textit{TrustVal}; the maximum (\textit{TrustValMax}) and minimum (\textit{TrustValMin}), and the trigger value (\textit{TrustTrig}) that, if \textit{TrustVal} went below it, \gls{csmrpl} will drop any \gls{rpl} control messages from that neighbor. the conceptual diagram of \textit{\gls{csmrpl}-Trust} interface is shown in Fig. \ref{fig_CSM-Trust}.

In addition, the calculation of \textit{TrustVal} is left to the external mechanism and can be dynamic to allow for larger flexibility to different applications and scenarios. For example, an \gls{ids} can use its own methods (e.g., special messages, monitoring the traffic, etc.) to determine the neighbor's trustworthiness, then it updates \textit{TrustVal} in the SC table, which tells \gls{csmrpl} if it should accept or reject control messages from that neighbor.

\section{Evaluation of the Chained Secure Mode}\label{CSMEval}
To evaluate our proposed \gls{csmrpl}, we conducted a comparison on security and performance between our devised prototype of \gls{csmrpl} and the currently implemented secure modes: \gls{rpl} in \gls{umrpl} (vanilla ContikiRPL), \gls{psmrpl}, and \gls{psmrpl}rp (both according to Perazzo \textit{et al.}\cite{Perazzo2017} implementation). All secure modes were evaluated against three routing attacks (\gls{na}, \gls{ca}, and \gls{wh}).
\squeezeup
\subsection{Evaluation Setup and Assumptions}\label{EvalSetup}
\begin{figure}[!t]
	\centering
	\includegraphics[scale=0.22]{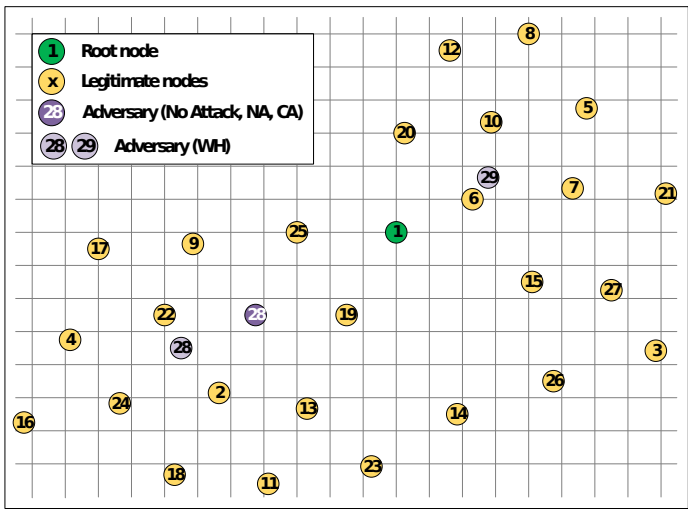}%
	\caption{Network topology used for the evaluation. The adversaries' locations are represented by a dark-purple circle (No Attack, \gls{na}, and \gls{ca} scenarios) or the light-purple ones (\gls{wh} scenario.)}
	\label{fig_Tplgy}
	\squeezeup
\end{figure}

The default simulator for Contiki \gls{os}\cite{ContikiOSRef}, Cooja, was used for all the simulations (with simulated motes). The topology used in our evaluation is shown in Fig. \ref{fig_Tplgy}, and simulation parameters are listed in Table \ref{table_1}.

As the chosen metrics for the evaluation, the average data \gls{pdr}, average data \gls{e2e} latency, and the average network power consumption per received data packet were all used for the comparison.

The following assumptions were used in our evaluation: For all the evaluated secure modes, the default \gls{of} for \gls{rpl} was used, i.e., \gls{mrhof}\cite{RFC6719}. Settings of Contiki \gls{os}'s uIP stack were as follows: IEEE 802.15.4\cite{STND802154} for the Physical layer and \gls{macl} sublayer, ContikiMAC\cite{ContikiMAC2011} and NullRDC\cite{ContikiOSRef} for the \gls{rdc} sublayer (see below), IPv6 and \gls{rpl} at the Network layer, and UDP for the Transport layer. To keep the focus on \gls{rpl}, we assumed neither security measures nor encryption were enabled at the Link layer. Data packets are sent toward the root by legitimate nodes only and at a rate of 1 packet/minutes per node.

As explained in our previous work\cite{Raoof2020}, Contiki \gls{os} supports several \gls{rdc} protocols\cite{ContikiOSRef}, with the ContikiMAC and NullRDC being the most common ones. The main difference between the two is that ContikiMAC is designed to aggressively conserve energy more than NullRDC, at the expense of having longer \gls{e2e} latency\cite{Raoof2020,Barnawi2019}. In this paper, the implementation of the \gls{wh} attack is based on our work in\cite{Raoof2020}; hence, it is only available using NullRDC protocol.

The data traffic model used for the evaluation, as described above, is a deterministic one that mimics a typical sensing-\gls{iot} network, where nodes send their sensor readings toward the root node at predetermined periods.

To test the external mechanism integration capability of \gls{csmrpl}, the following simple, proof-of-concept, external security mechanism was implemented (only in \gls{csmrpl} experiment):
\begin{itemize}
	\item \textit{TrustValMin}, \textit{TrustValMax}, and \textit{TrustTrig} were set to 0, 100, and 50, respectively.
	\item For the first \gls{rpl} message from a neighbor, a successful reception will set \textit{TrustVal} to \textit{TrustValMax}.
	\item Afterward, \textit{TrustVal} will increase or decrease based on the successful (or unsuccessful) decoding of the received \gls{rpl} control messages. The increment/decrement amount was set to 10.
\end{itemize}

The results obtained from the simulations were averaged over ten rounds per experiment with a 95\% confidence level.
\begin{table}[!t]
	\caption{List of Simulation Parameters (per \gls{rdc} protocol)}
	\label{table_1}
	\begin{tabular}{ll}
		\toprule
		\textbf{Description}&\textbf{Value}\\
		\midrule
		No. of simulation sets&
		\begin{minipage}{0.48\columnwidth}
			Two: one for each adversary type (\textbf{\underline{I}}nternal and \textbf{\underline{E}}xternal) (See \S\ref{AttMdl})
		\end{minipage} \\
		\midrule
		No. of experiments per set&
		\begin{minipage}{0.48\columnwidth}
			Four: one for each secure mode (\gls{umrpl}, \gls{psmrpl}, \gls{psmrpl}rp, and \gls{csmrpl})
		\end{minipage} \\
		\midrule
		No. of scenarios per experiment&
		\begin{minipage}{0.48\columnwidth}
			3 (ContikiMAC) / 4 (NullRDC) - See \S\ref{EvalSetup}
		\end{minipage} \\
		\midrule
		Sim. rounds per scenario / time&10 rounds / 20 min. per round\\
		\midrule
		Node positioning&Random distribution\\
		\midrule
		Deployment area&210m W x 150m L\\
		\midrule
		Number of nodes&
		\begin{minipage}{0.49\columnwidth}
			28 (29 for the \gls{wh} scenario)\\includes 1 adversary (2 for \gls{wh})
		\end{minipage} \\
		\midrule
		Sensor nodes type&
		\begin{minipage}{0.48\columnwidth}
			Arago Sys. Wismote mote
		\end{minipage}\\
		\bottomrule
	\end{tabular}
\end{table}
\squeezeup
\subsection{Adversary Model and Attack Scenarios}\label{AttMdl}
The following attacks were chosen due to the low cost for the adversary to launch them, as they require little or no processing of \gls{rpl}’s messages. At the same time, the effect of these attacks can be significant on the network. The location of the adversary(ies) was chosen to present the most prominent effect of the investigated attacks\cite{Pu2018a,Wallgren2013,Mayzaud2016}.
\begin{figure*}[!ht]
	\centering
	\subfloat[]{\includegraphics[height=4.7cm, width=.30\linewidth]{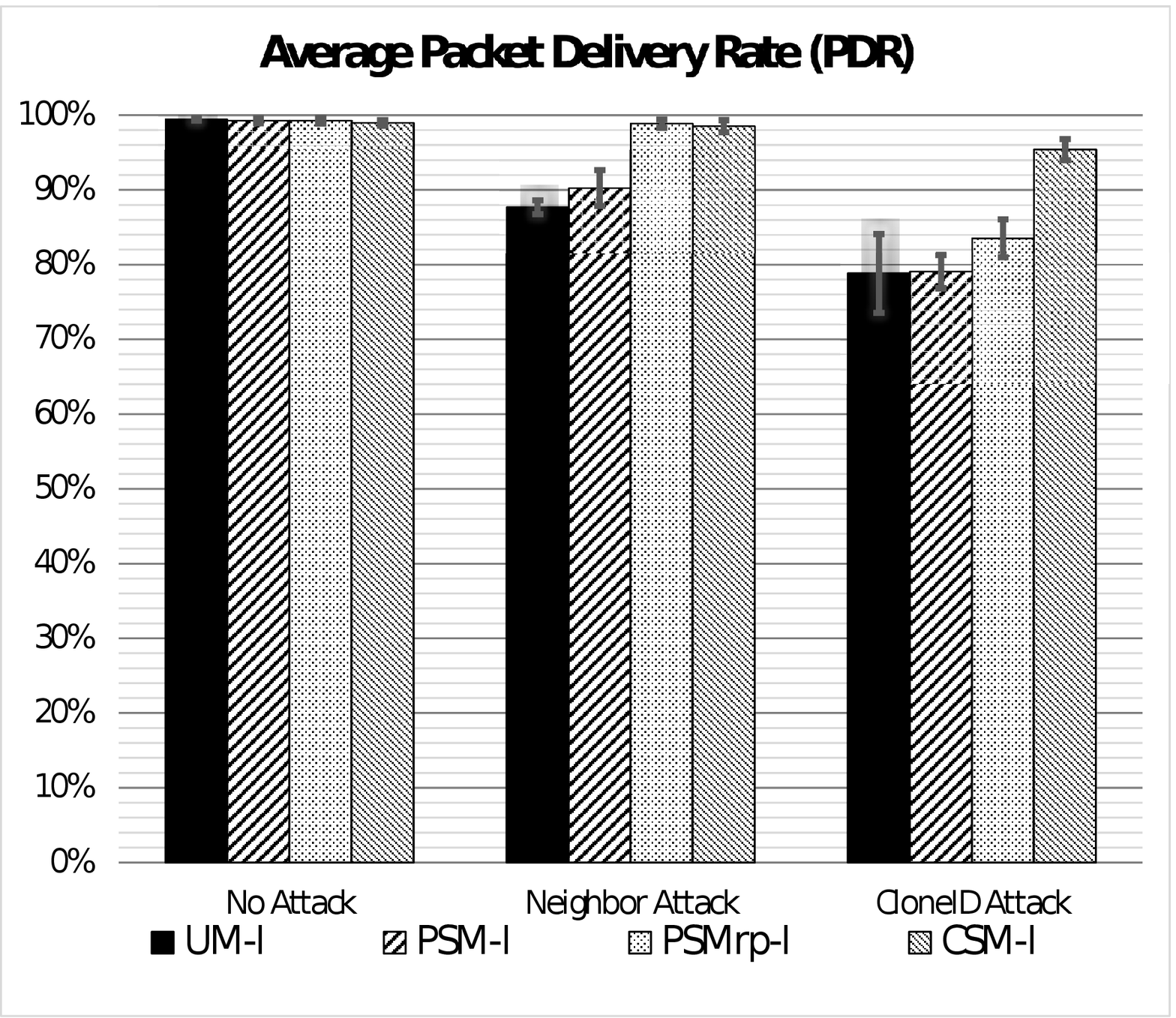}%
		\label{fig_PDR-CMAC-I}}
	\hfil
	\subfloat[]{\includegraphics[height=4.7cm, width=.30\linewidth]{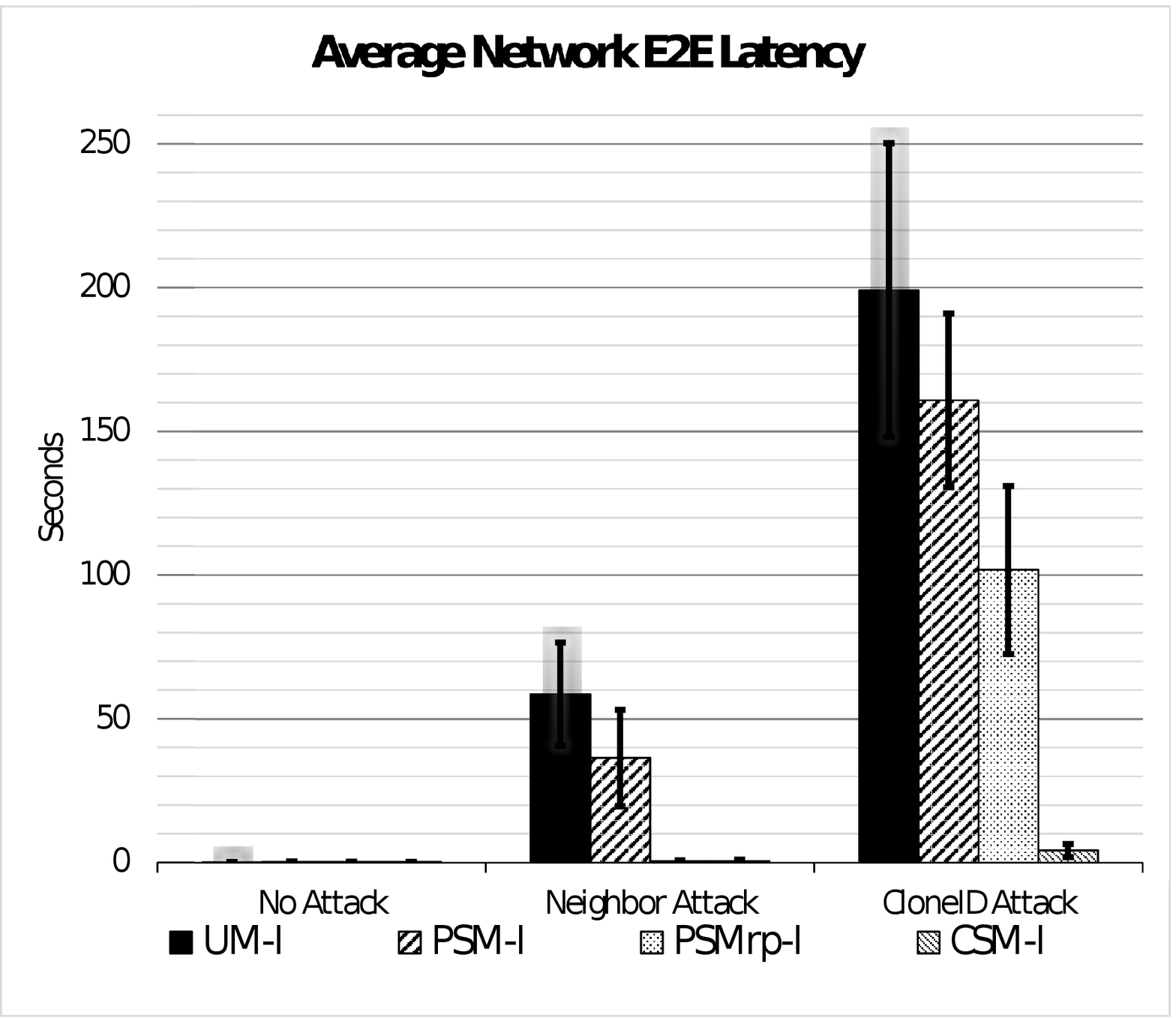}%
		\label{fig_E2E-CMAC-I}}
	\hfil
	\subfloat[]{\includegraphics[height=4.7cm, width=.363\linewidth]{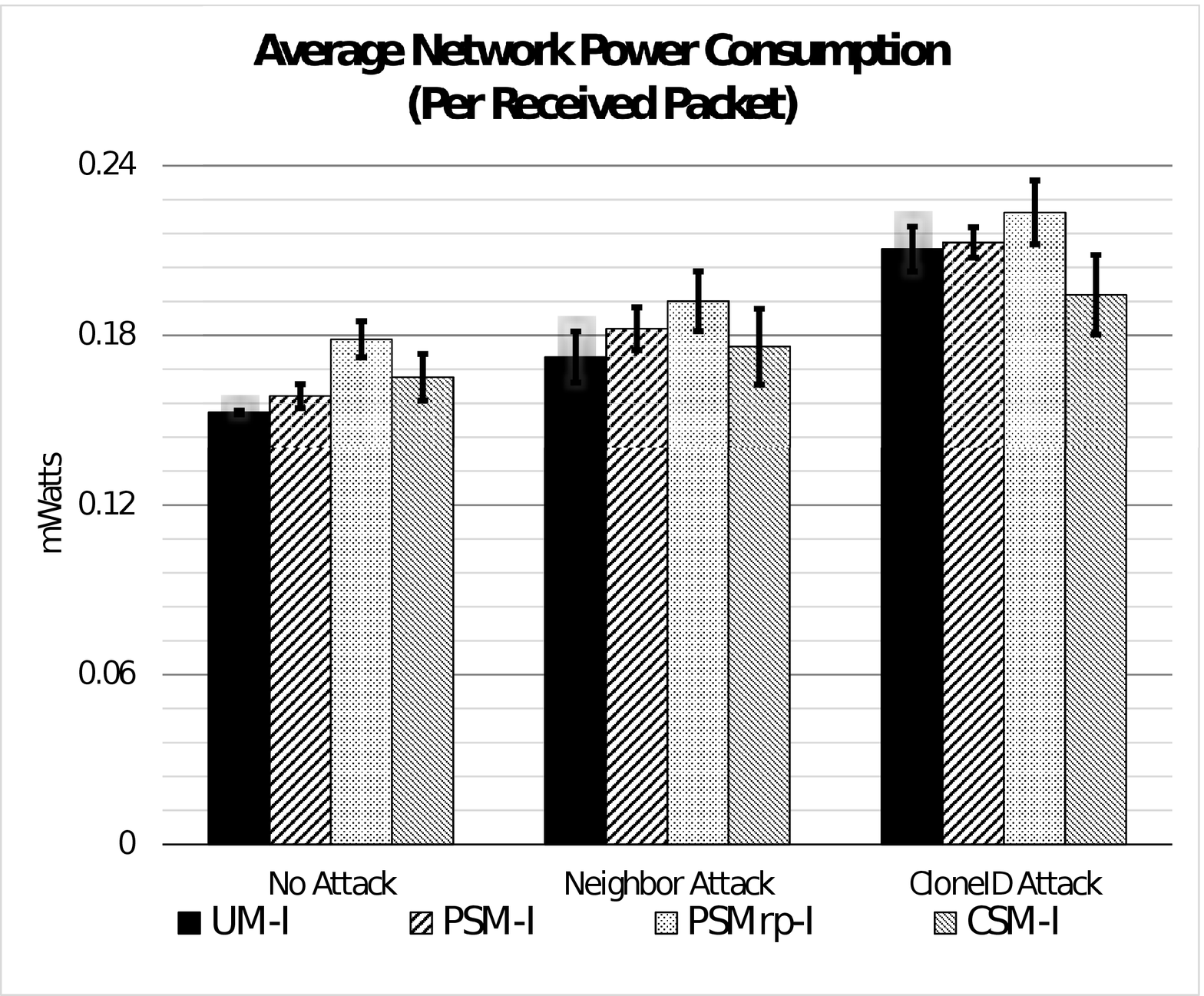}%
		\label{fig_PWR-CMAC-I}}
	\hfil
	\caption{Simulation results for the four experiments (three attacks scenarios - internal adversary), using ContikiMAC RDC protocol.}%
	\label{fig_Rslts-CMAC-I}
	\squeezeup
\end{figure*}
\begin{figure*}[!ht]
	\centering
	\subfloat[]{\includegraphics[height=4.7cm, width=.30\linewidth]{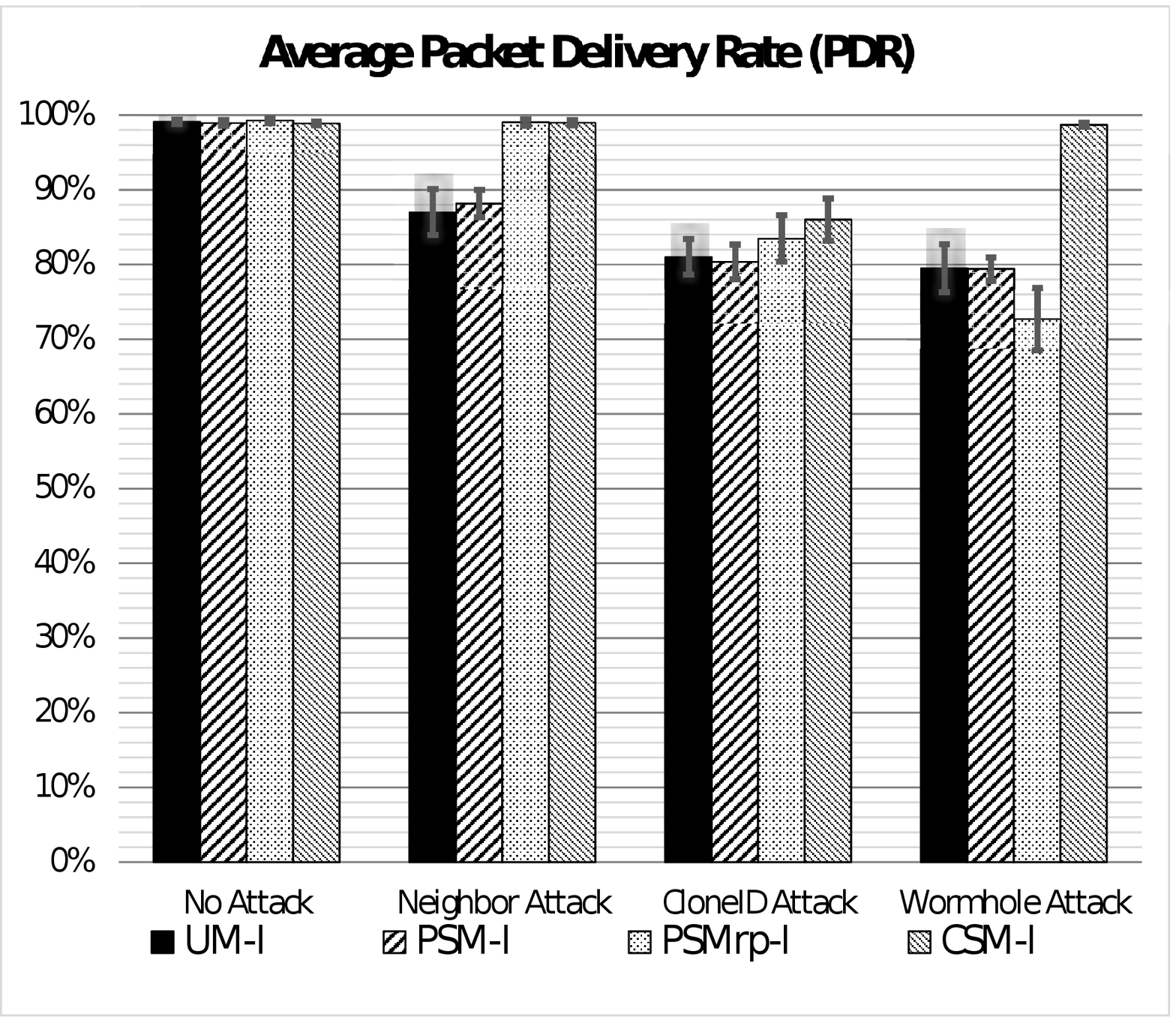}%
		\label{fig_PDR-NR-I}}
	\hfil
	\subfloat[]{\includegraphics[height=4.7cm, width=.30\linewidth]{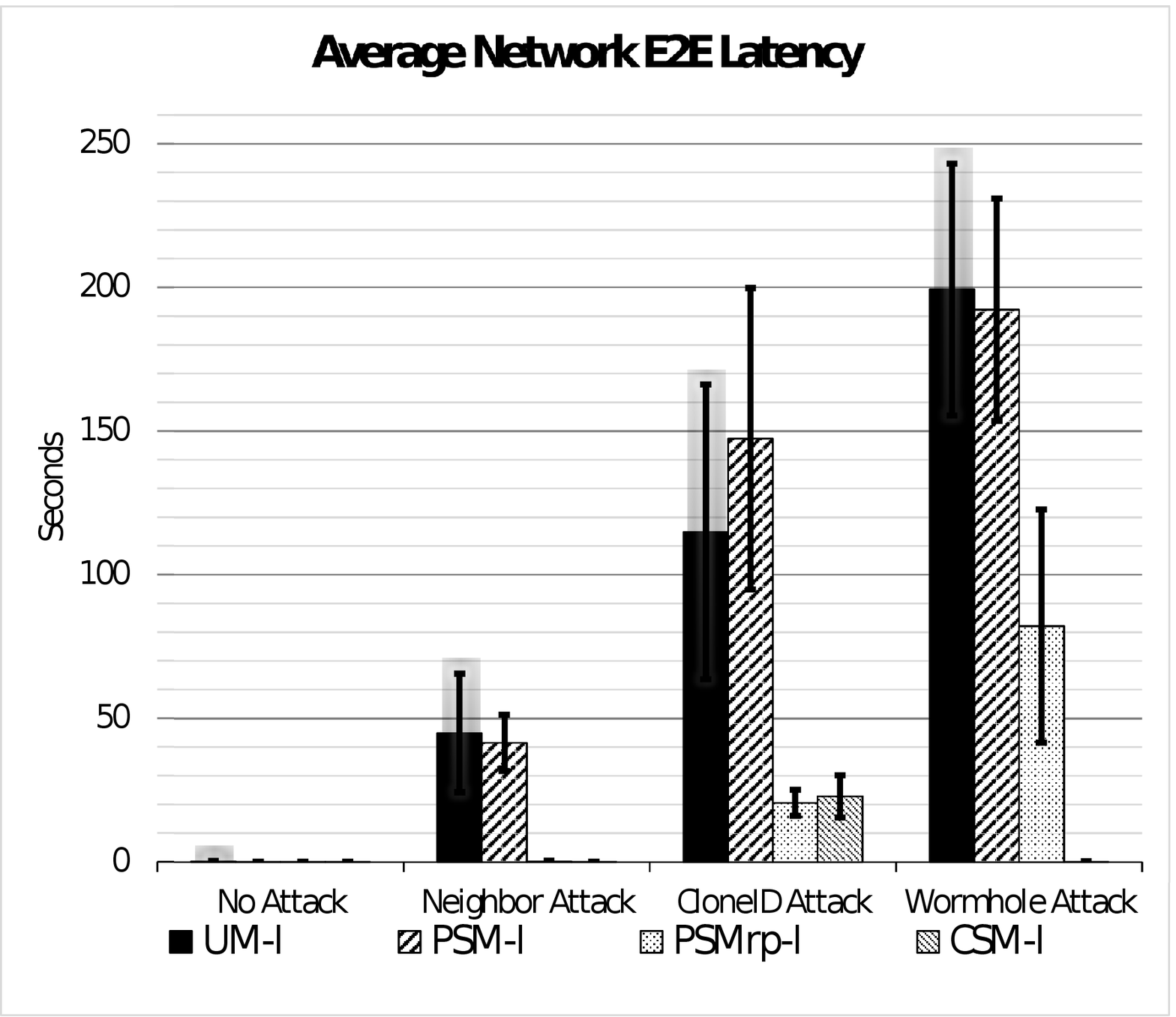}%
		\label{fig_E2E-NR-I}}
	\hfil
	\subfloat[]{\includegraphics[height=4.7cm, width=.363\linewidth]{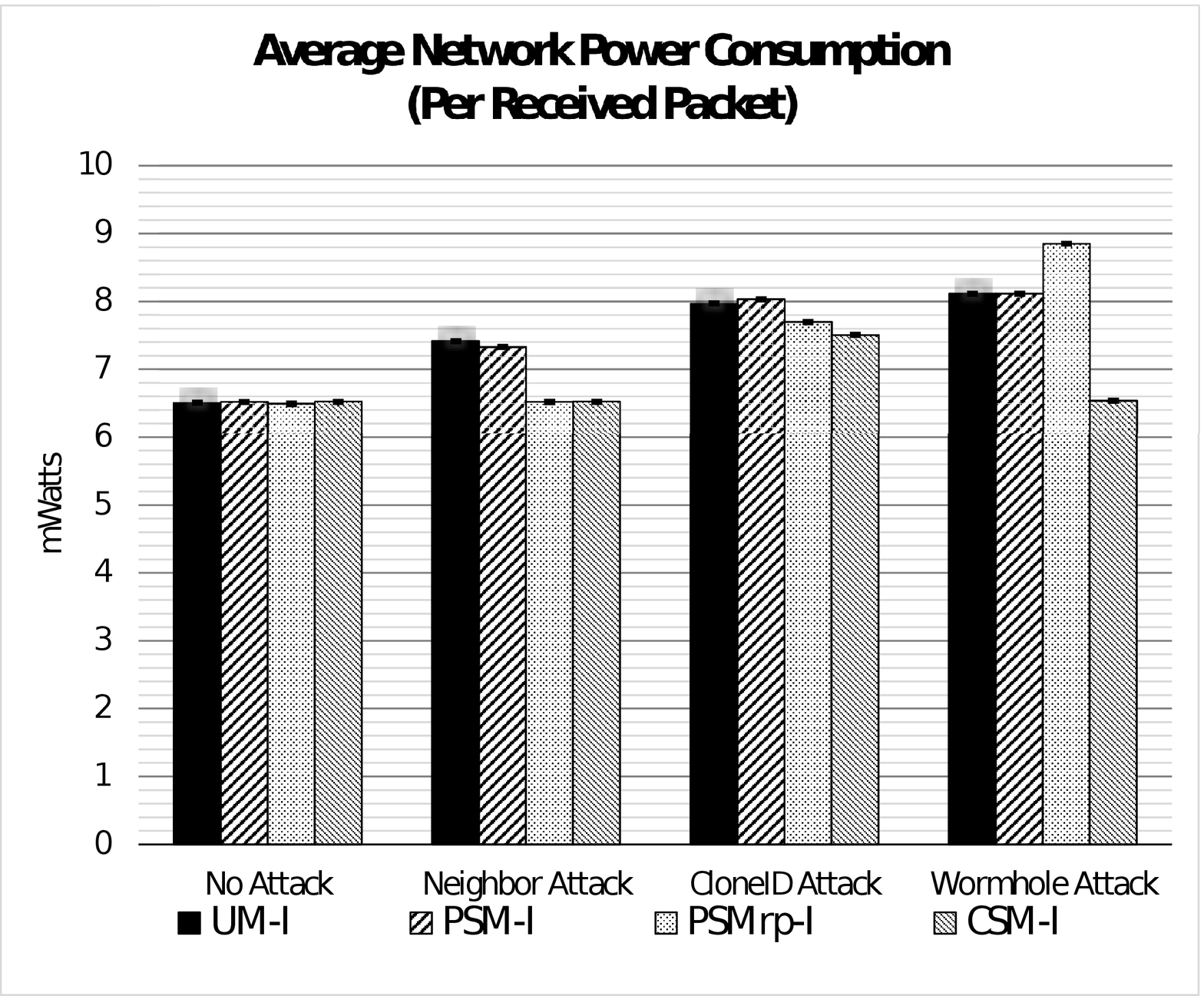}%
		\label{fig_PWR-NR-I}}
	\hfil
	\caption{Simulation results for the four experiments (four attacks scenarios - internal adversary), using NullRDC RDC protocol.}
	\label{fig_Rslts-NR-I}
	\squeezeup
\end{figure*}

All the secure modes were evaluated in both normal operation (\textit{No Attack}) and against three replay routing attacks\cite{Raoof2018,Wallgren2013} (the following depicts our implementation of the attacks):-
\begin{enumerate}
	\item \textbf{\acrfull{na}:} Whenever the adversary hears a \gls{dio} message from any neighbor (regardless of its destination), it will replay it (as a multicast) to all its neighbors without modifications or processing, deluding them to think that the original sender is within their range. If the original sender has a better rank (see \S\ref{background}), the receivers of the replayed message will select it as their preferred parent, which may result in lost data packets and longer \gls{e2e} delays\cite{Raoof2020}. The objectives of this attack\cite{Raoof2018} are the disruption of data packets transmission and the disconnection of the routing topology.
	\item \textbf{\acrfull{ca}:} The adversary will clone the identity of another node, in our case node (25), by changing its \gls{ipv6} and \gls{macl} addresses to match that of the cloned node (by monitoring its frames and \gls{rpl} messages). In addition, it will follow and copy the cloned node's rank by reading its \gls{dio} messages. Our implementation combined \gls{ca} with a \gls{sf} attack (that only drops data packets passing through the adversary) to better show how the \gls{ca} changed the \gls{dodag} around the adversary. The main goals of this attack\cite{Raoof2018} are to disrupt the routing topology and manipulate any reputation-based \gls{ids}.
	\item \textbf{Out-of-Band \acrfull{wh} attack:} Two adversaries (connected by an out-of-band link) will forward and replay all \gls{rpl} control messages they hear from their neighbors between the two locations where they reside. Due to simulation limitations\cite{Raoof2020}, this attack scenario is only available in the NullRDC set of experiments. The attack aims\cite{Raoof2018} to increase data packets' latency, disrupt the routing topology, exhaust the victim nodes' energy, and disconnect parts of the network.
\end{enumerate}

For the adversaries, they run in the same \gls{rpl} secure mode as the legitimate nodes. However, they have two types: \textit{Internal} adversaries, where they have the proper preinstalled encryption key for \gls{psmrpl}, \gls{psmrpl}rp, and \gls{csmrpl} experiments; and the \textit{External} adversaries, which do not have the required encryption key. Also, it is worth mentioning that the external and internal versions of the adversaries in \gls{umrpl} scenario are the same.

In all cases, the adversary starts as a legitimate node, tries to join the network, then launches the attack after two minutes. For the Wormhole attack, the two adversaries are always in promiscuous mode and never participate in the \gls{dodag}.

\section{Results for Internal Adversary Sets}\label{INTResults}%\squeezeupBY{2}
\subsection{Effects on the Data \gls{pdr}}%\squeezeupBY{1.5}
Looking at Fig. \ref{fig_PDR-CMAC-I} (ContikiMAC) and Fig. \ref{fig_PDR-NR-I} (NullRDC), it is clear that \gls{psmrpl}rp and \gls{csmrpl} (for both ContikiMAC and NullRDC) successfully eliminated the \gls{na} effect, with both of them having almost 100\% \gls{pdr}. \gls{umrpl} and \gls{psmrpl} suffered more (\gls{pdr}$\approx$ 80-90\%) as the adversary actually was able to become part of the network.

For the \gls{ca} scenario, it is noticeable that the attack was able to confuse the surrounding nodes and in many cases it successfully switched their preferred parent to the adversary for the \gls{umrpl}, \gls{psmrpl}, and \gls{psmrpl}rp. This shows in the lower \gls{pdr} (\gls{pdr}$\approx$ 80-85\%) for all secure modes except \gls{csmrpl}. On the other hand, \gls{csmrpl} was able to reduce the effect of the attack (\gls{pdr}$\approx$ 85-95\%), albeit the reduction here is caused by \gls{csmrpl}'s integration with the external security mechanism, which acted as a dynamic blacklist and "\textit{untrusted}" both the legitimate and cloned nodes after several unsuccessful recovery attempts.
%\begin{samepage}
	\begin{figure*}[!ht]
		\centering
		\subfloat[]{\includegraphics[height=4.7cm, width=.30\linewidth]{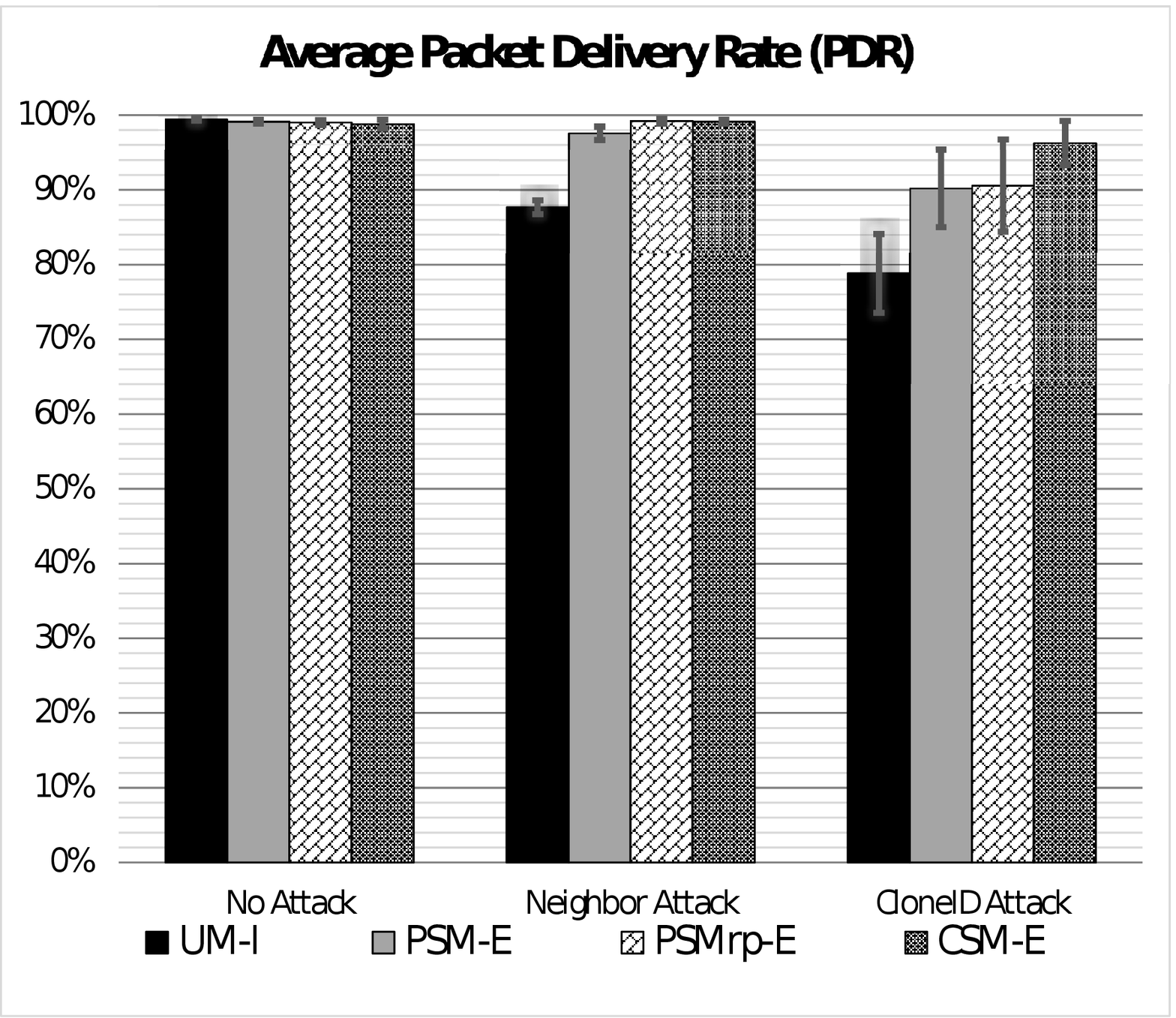}%
			\label{fig_PDR-CMAC-E}}
		\hfil
		\subfloat[]{\includegraphics[height=4.7cm, width=.30\linewidth]{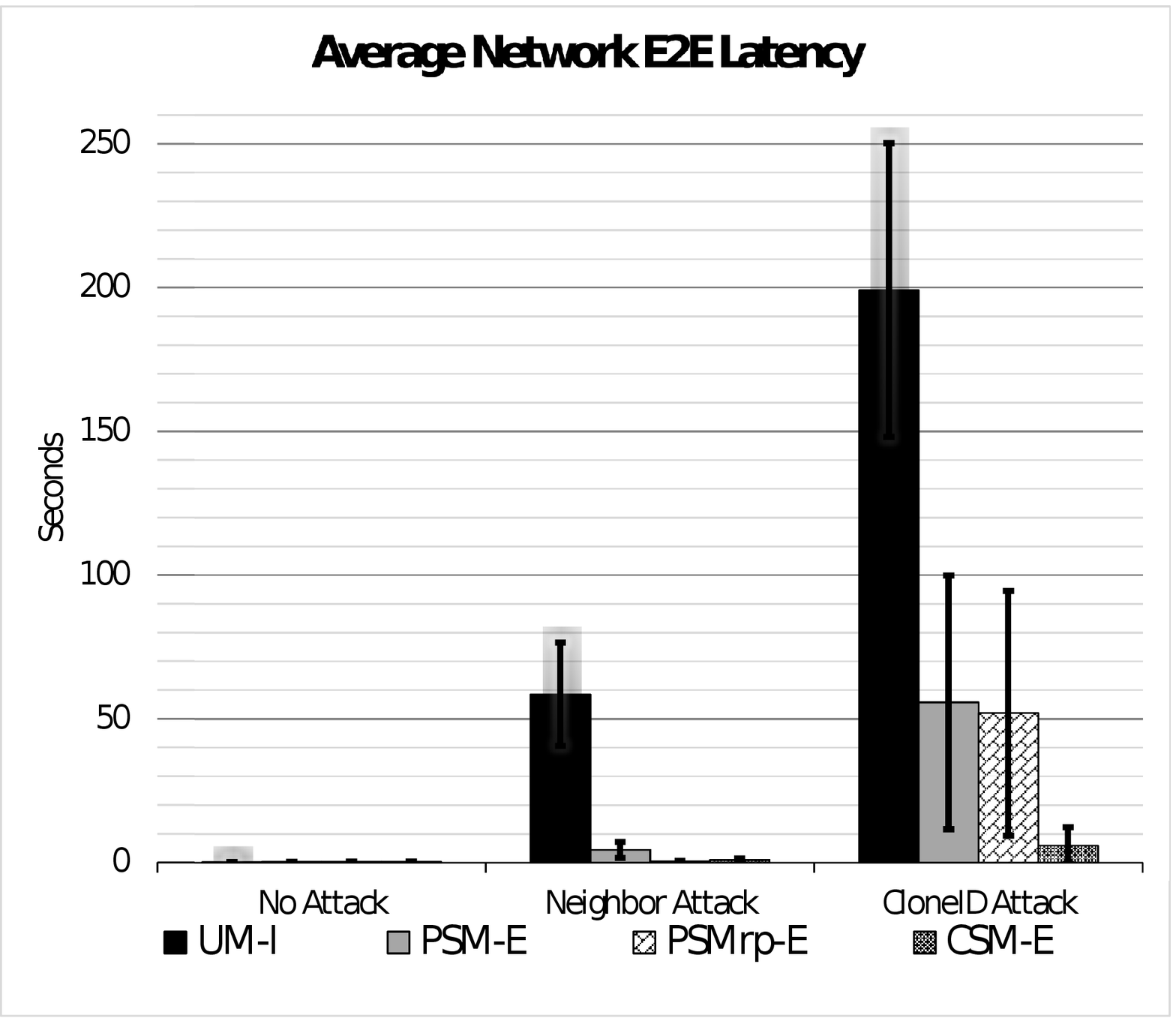}%
			\label{fig_E2E-CMAC-E}}
		\hfil
		\subfloat[]{\includegraphics[height=4.7cm, width=.363\linewidth]{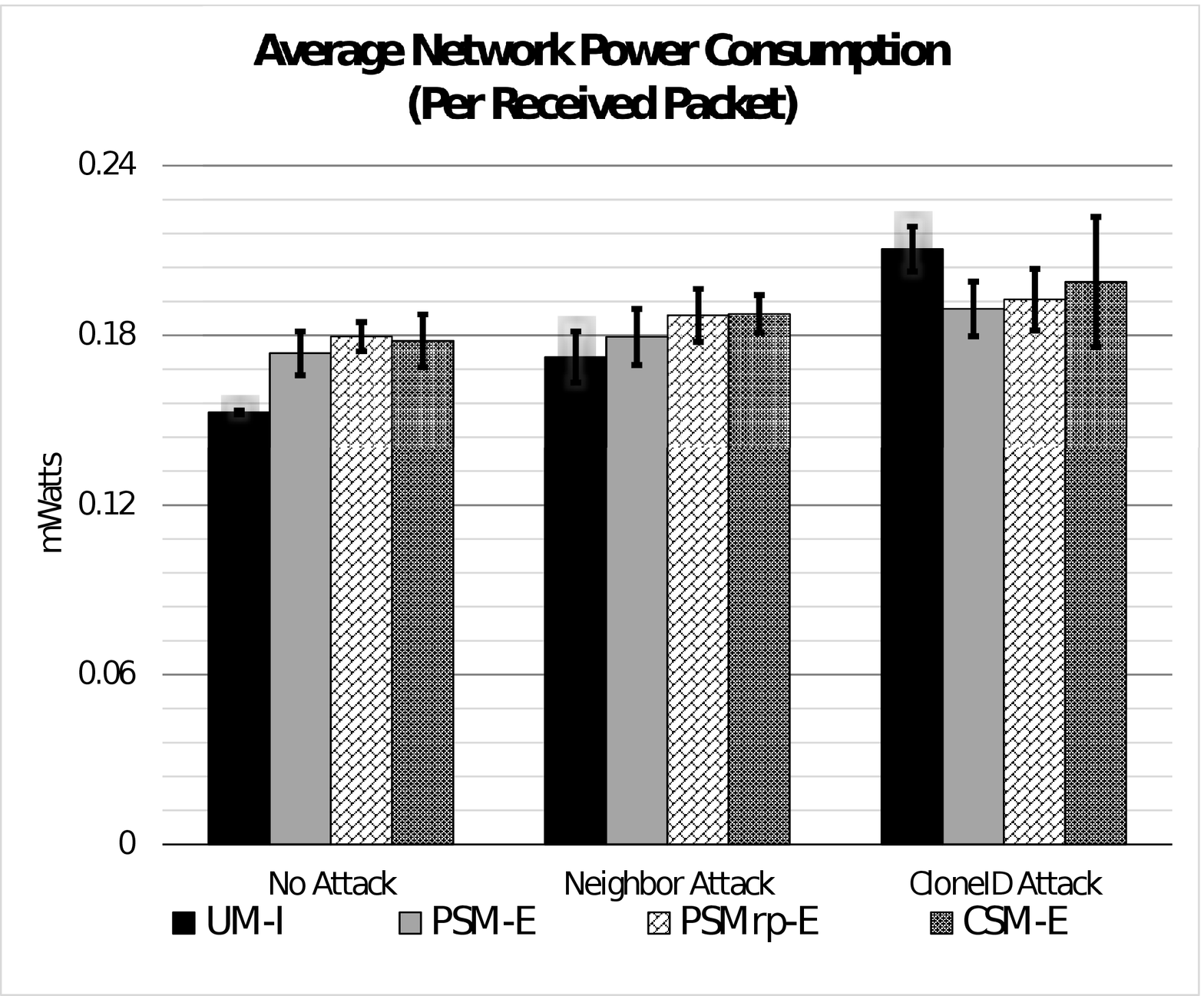}%
			\label{fig_PWR-CMAC-E}}
		\hfil
		\caption{Simulation results for the four experiments (three attacks scenarios - external adversary), using ContikiMAC RDC protocol.}
		\label{fig_Rslts-CMAC-E}
		\squeezeup
	\end{figure*}
%	\nopagebreak
	\begin{figure*}[!ht]
		\centering
		\subfloat[]{\includegraphics[height=4.7cm, width=.30\linewidth]{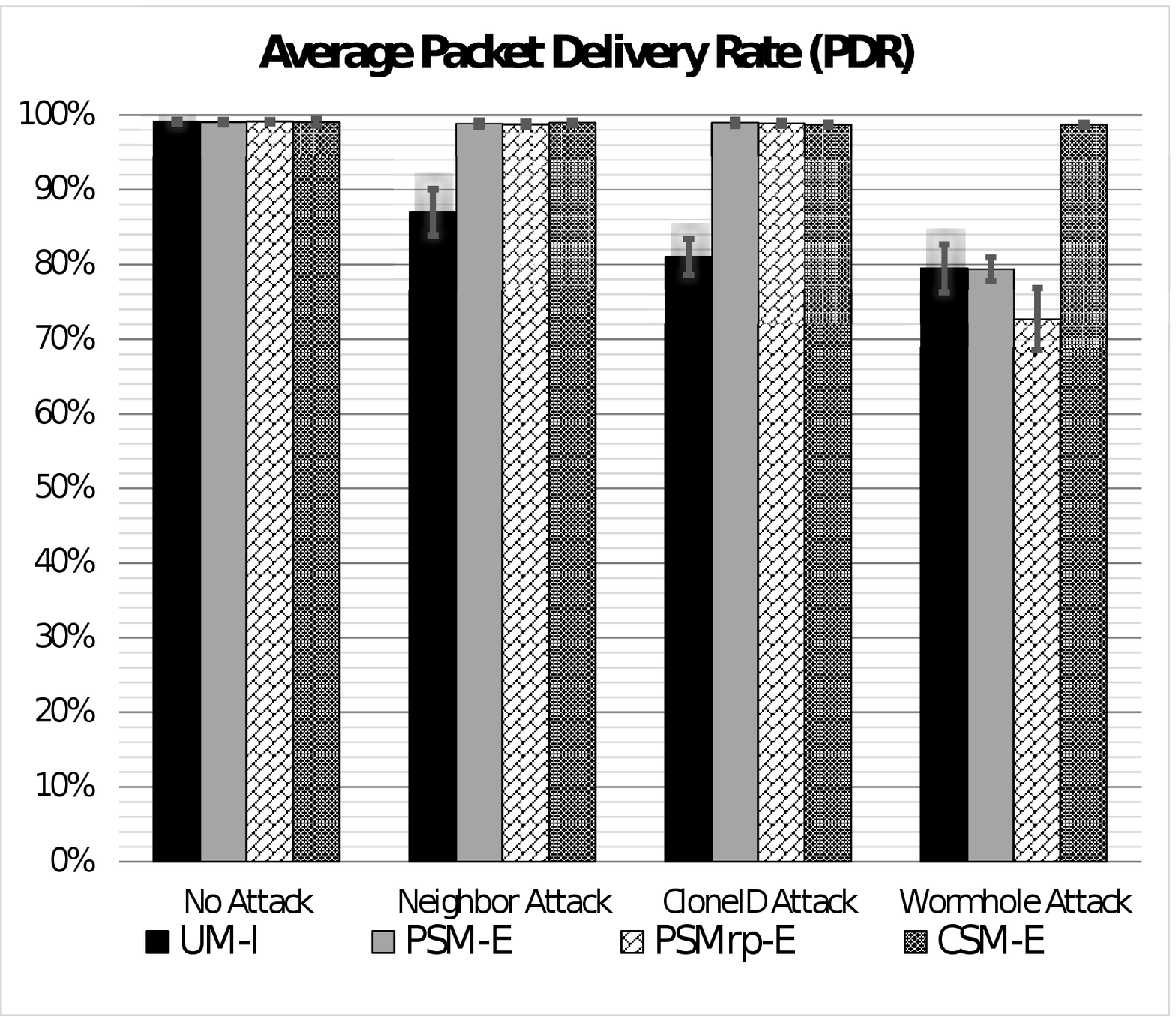}%
			\label{fig_PDR-NR-E}}
		\hfil
		\subfloat[]{\includegraphics[height=4.7cm, width=.30\linewidth]{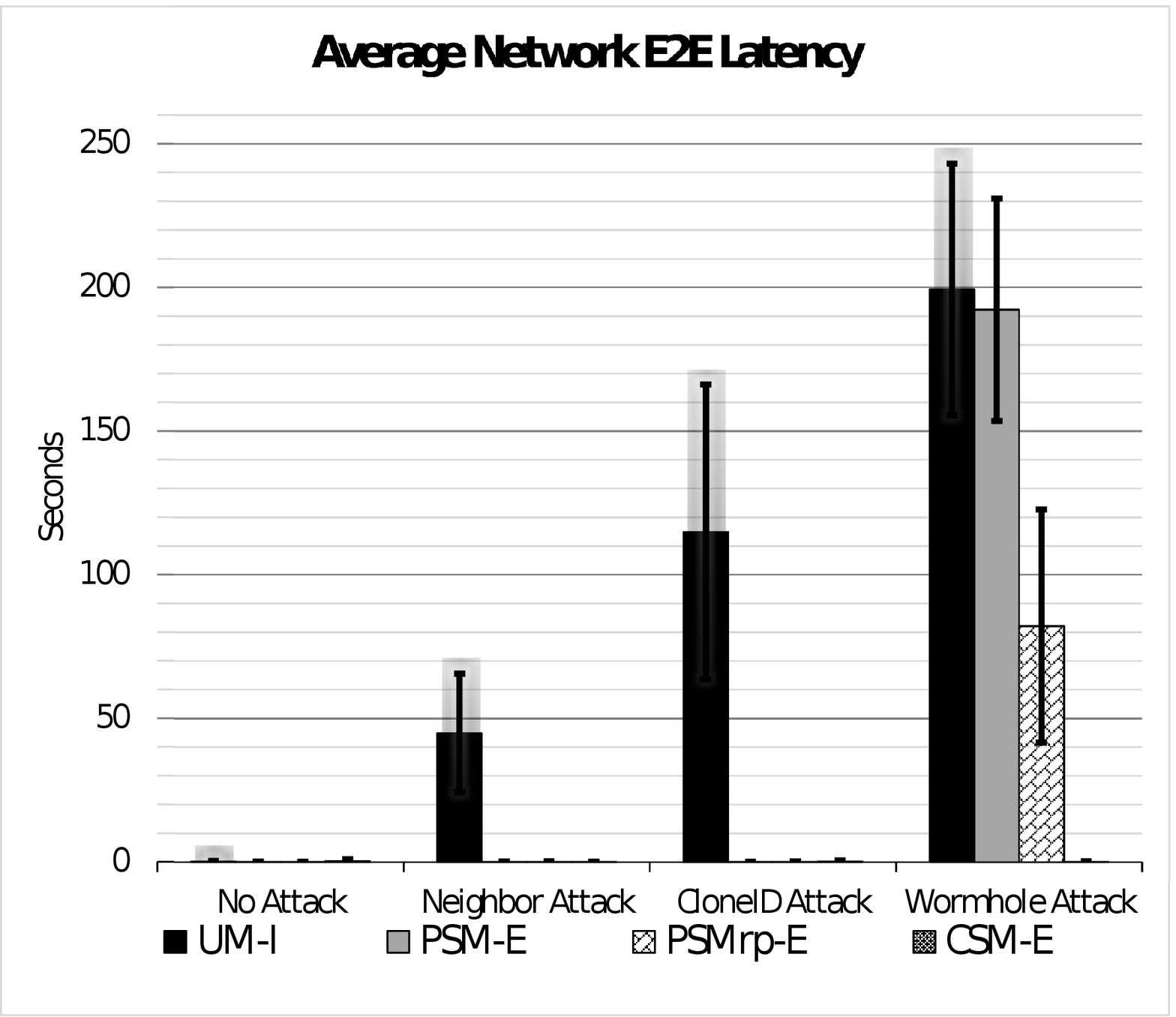}%
			\label{fig_E2E-NR-E}}
		\hfil
		\subfloat[]{\includegraphics[height=4.7cm, width=.363\linewidth]{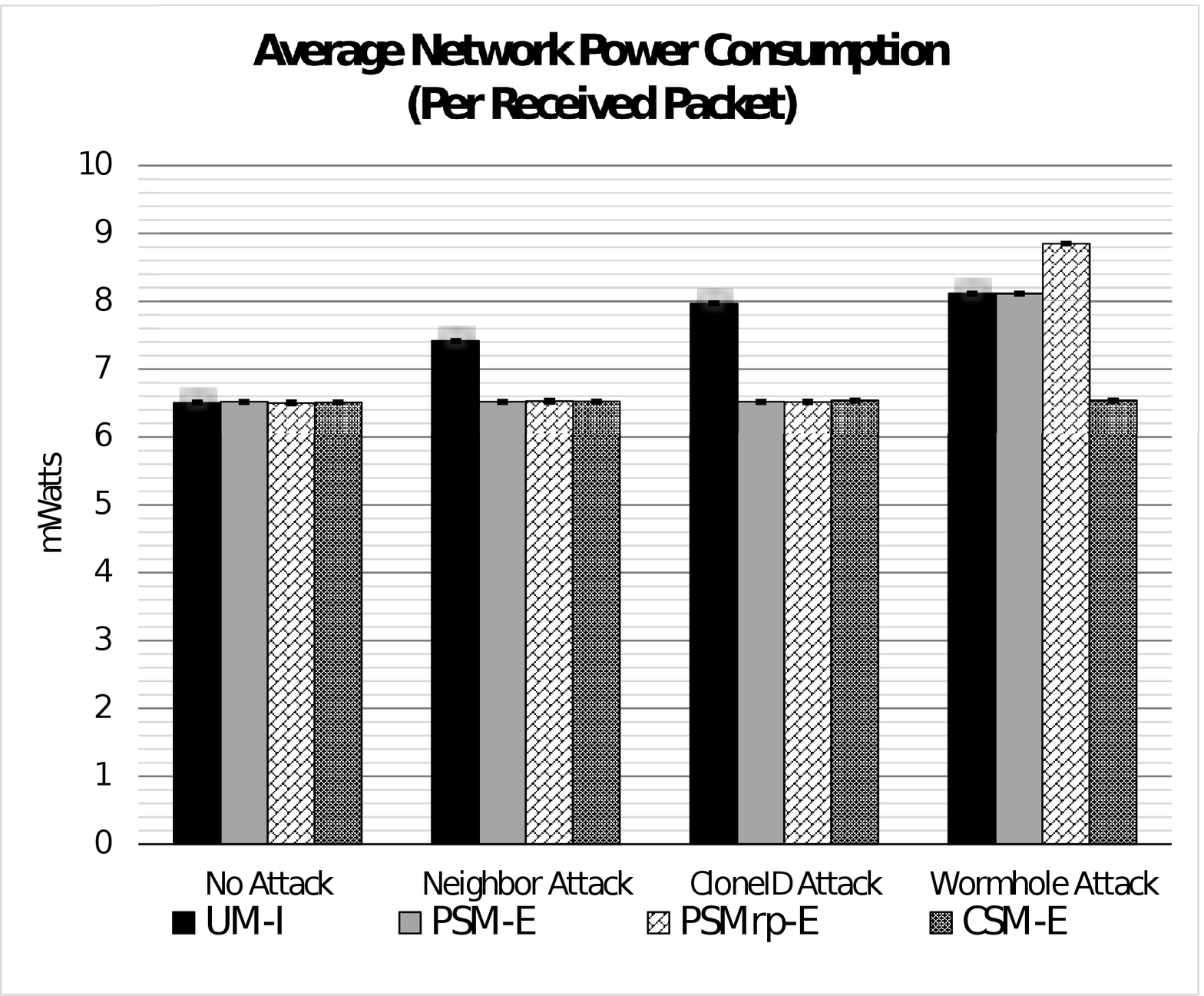}%
			\label{fig_PWR-NR-E}}
		\hfil
		\caption{Simulation results for the four experiments (four attacks scenarios - external adversary), using NulRDC RDC protocol.}
		\label{fig_Rslts-NR-E}
		\squeezeup
	\end{figure*}
%\end{samepage}
%\squeezeup

Finally, \gls{csmrpl} outperformed the other secure modes in the \gls{wh} attack scenario, where it was able to mitigate the attack (\gls{pdr}$\approx$ 95-99\% compared to 75-80\% for the other secure modes). This is mainly due to \gls{csmrpl} policy of dropping, without processing, \gls{rpl} control messages from new neighbors if they were encoded with unknown \gls{sc} values. Hence, the routing topology will not change due to the \gls{wh} attack.

\subsection{Effects on the Data \gls{e2e} Latency}
Looking at Fig. \ref{fig_E2E-CMAC-I} (ContikiMAC) and Fig. \ref{fig_E2E-NR-I} (NullRDC), it can be seen that \gls{na} has been mitigated by both \gls{psmrpl}rp and \gls{csmrpl} (latency $\approx$ a few milliseconds), and that it introduced higher \gls{e2e} latency to the network for the other secure modes ($\approx$40-70 seconds), confirming our previous findings in \cite{Raoof2019a, Raoof2020}.

The results for the \gls{ca} shows the significant effect of the attack on the network, where the latency is in the (100-200) seconds range for \gls{umrpl} and \gls{psmrpl}, while \gls{psmrpl}rp varies between 20 seconds (NullRDC) to 100 seconds (ContikiMAC). This variation is caused by ContikiMAC as its energy-conservative mechanism amplifies any latencies due to transceiver extended \textit{sleep} times\cite{Raoof2020, Barnawi2019}.

Further proofing \gls{csmrpl}'s ability to mitigate the \gls{wh} attack, latency is kept to minimum (<5 seconds) compared to the other secure modes ($\approx$80-200 seconds), for both \gls{rdc} protocols.

\subsection{Effects on Power Consumption}
Comparing Fig. \ref{fig_PWR-CMAC-I} (ContikiMAC) and \ref{fig_PWR-NR-I} (NullRDC), it can be seen that all secure modes have similar patterns, where the \gls{ca} scenario has the higher power consumption in the ContikiMAC set, and the \gls{wh} attack scenario has the highest power consumption in the NullRDC set. However, \gls{csmrpl} has the lowest power consumption in every situation when compared to other secure modes. This is more prominent in the \gls{wh} attack scenario.  It is worth mentioning that power readings for NullRDC are higher than the ones for ContikiMAC, as the transceiver is always on for the former while it is off most of the time for the latter\cite{Raoof2019a, Raoof2020}. 
\squeezeup
\section{Results for External Adversary Sets}\label{EXTResults}
\subsection{Effects on the Data \gls{pdr}}
Fig. \ref{fig_PDR-CMAC-E} (ContikiMAC) and Fig. \ref{fig_PDR-NR-E} (NullRDC) show that \gls{psmrpl}, \gls{psmrpl}rp, and \gls{csmrpl} are capable of mitigating both \gls{na} and \gls{ca}, with the latter having a slightly more effect in the case of ContikiMAC set (\gls{pdr} $\approx$ 90-99\%). However, \gls{csmrpl} was the only secure mode that mitigated the \gls{wh} attack with (\gls{pdr} $\approx$ 98\%) compared to ($\approx$ 70-75\%) for the other secure modes.

To explain the effect of the \gls{ca} attack on the network when ContikiMAC is used, we must first understand the way \gls{rpl} decides on selecting the preferred parent or switching to another parent. In general, \gls{rpl} depends on neighbors' statistics, which are provided by \gls{rpl} itself, IPv6 Network Discovery protocol\cite{RFC6775}, and Link-layer protocols. In ContikiMAC, the Link-layer statistics depends on the successfully-received frames, probing frames, and acknowledgment frames from these neighbors, among other factors\cite{Uwase2017}. Now, when \gls{rpl} operates on top of ContikiMAC, it will use these statistics, besides its own, to decide if the neighbor is still "\textit{fresh}" or not; hence, if it should keep it in its possible-parents list or even as the preferred parent\cite{RFC6550}. However, NullRDC does not provide such statistics, leaving \gls{rpl} only to use its statistics.

Since the dropped \gls{rpl} control messages are still successfully received by ContikiMAC, it will affect \gls{rpl}'s decision on keeping the victim node (and the cloned node - adversary) as the preferred parent at the neighboring nodes. In many cases, it will prolong the time required to switch to another parent until \gls{rpl}'s statistics point to a communication failure and force the switch, unlike the case for NullRDC where \gls{rpl} quickly detects the communication failure and switch to another parent.
\squeezeupBY{2.5}
\subsection{Effects on the Data \gls{e2e} Latency}
similar to the analysis of \gls{pdr}, Fig. \ref{fig_E2E-CMAC-E} (ContikiMAC) and Fig. \ref{fig_E2E-NR-E} (NullRDC) show that the three secure modes have successfully mitigated the \gls{na}, while the \gls{ca} still has an effect in the case of ContikiMAC for the same reason mentioned above. Again, \gls{csmrpl} shows as the only secure mode capable of mitigating the \gls{wh} attack, with minimum \gls{e2e} latency.
\squeezeupBY{2.5}
\subsection{Effects on Power Consumption}
The results for ContikiMAC (Fig. \ref{fig_PWR-CMAC-E}) shows similar patterns among all the secure modes and for all attacks as a proof of significant attacks' mitigation. However, \gls{psmrpl}, \gls{psmrpl}rp, and \gls{csmrpl} do have a slightly more power consumption than \gls{umrpl} in the No Attack scenario, due to the additional security measures. This case is reversed for the \gls{ca} attack scenario, as the \gls{umrpl} was the only mode fully affected by the attack.

Fig. \ref{fig_PWR-NR-E} show that all three secure modes have similar patterns for the \gls{na} and \gls{ca} scenarios (significant mitigation of the attacks) under NullRDC. While, for the \gls{wh} attack, \gls{csmrpl} has the lowest power consumption (matching the No Attack scenario), due to the mitigation of the attack. 
\squeezeup
\section{Discussion}\label{Obsr}
Our observations from the evaluation experiments can be summarized in the following points:
\subsection{Enhanced Security Features of \gls{csmrpl}}
Those can be summarized as follows:
\begin{enumerate}[label=\roman*)]
	\item \gls{csmrpl} adds an extra layer of security by encoding the control messages and chaining them with the \gls{sc} values, providing a means of sender authentication which limits the adversaries' ability to eavesdrop on, manipulate, forge, and replay \gls{rpl} control messages.
	\item Encoding of the \textit{Code} field of the \gls{icmp} header in \gls{csmrpl} means that external adversaries cannot identify the type of \gls{rpl} control messages by reading the \gls{icmp} header, except for the first message of each message flow as it is currently encoded with zero - see Fig. \ref{fig_CSM}c. Hence, external replay attacks that target specific \gls{rpl} control messages (e.g., \gls{na}) can be mitigated by using \gls{csmrpl}.
	\item Unlike \gls{psmrpl}rp, \gls{csmrpl} provides mitigation to both "one-way" and "two-way replay attacks (e.g., the \gls{na} and \gls{wh}, respectively), due to the chaining of the control messages (by the \gls{sc} values) that acts as a sender authentication mechanism, without the need for a challenge/response mechanism as in \gls{psmrpl}rp.
\end{enumerate}

Due to the characteristics of the intra-flow \gls{nc}, there is one case where an internal adversary can launch one of the investigated attacks, which is if it is able to extract and track all the \gls{sc} values from the exchanged \gls{rpl} control messages between the adversary's neighbors, which requires the adversary to be around the victim nodes when the network starts operating. However, tracking all the \gls{sc} values for all the neighboring nodes' communications (and in other parts of the network for a \gls{wh} attack) would require a tremendous amount of resources (e.g., processing power, memory/storage capacity, and fast transceiver) to be available to the adversary. The amount of the required resources depends on the \gls{iot} network size, the location of the adversary, the number of victim nodes, and the type of attack desired.

\subsection{\gls{csmrpl} Reduction of the In-threat Period}
The \underline{in-threat period} for a replay-attack adversary can be defined as "\textit{the time duration in which an adversary can overhear and understand the whole (or a part of) the exchanged \gls{rpl} control messages and launch replay attacks}". This period ranges between \textbf{zero} (\textit{the adversary cannot launch attacks successfully}) to \textbf{infinity} (\textit{the adversary can launch attacks at any time}), depending on the secure mode used, the adversary type, and the attack.

For \gls{umrpl}, the in-threat period is always \textbf{infinity} as the adversary can understand \gls{rpl} messages and launch attacks at any time. On the other hand, the in-threat period for \gls{psmrpl} can be either:
\begin{itemize}
	\item \textbf{Infinity} for all internal adversaries, or external adversaries of replay/identity-cloning attacks. The former can decrypt the whole control message with the preinstalled encryption key at any time, while the latter can identify \gls{rpl} control messages from the "Type" and "Code" fields of the \gls{icmp} header, then replay them at any other time without the need to decrypt the message contents.
	\item \textbf{Zero} for external adversaries of attacks that require a full understanding of \gls{rpl} control messages, due to the lack of the used encryption key.
\end{itemize} 

As \gls{csmrpl} enhances \gls{rpl} security through the intra-flow \gls{nc}, it limits the adversaries' ability to launch several internal and external attacks that require identifying and understanding \gls{rpl} control messages. Hence, \gls{csmrpl} significantly reduces the in-threat period to either:
\begin{itemize}
	\item \textbf{The time between receiving the first \gls{mc} and the first \gls{uc} messages} for all internal adversaries. During this period, the adversary will try to intercept the first \gls{uc} control message (which is encoded with zeros and has the \gls{sc} values for both \gls{uc} and \gls{mc} flows), so it can use the included \gls{sc} values to decode (then decrypt) the following message from any flow. However, the adversary needs to continuously intercept and decode all messages from its victims in order to keep up with used \gls{sc} values, which significantly raises the cost of any attack's launch.
	\item \textbf{Zero} for all external adversaries, due to the lack of both the used encryption key and the correct \gls{sc} values. In addition, \gls{csmrpl}'s encoding of the "Code" field of the \gls{icmp} header makes it harder to the adversary to identify the type of \gls{rpl} control message; hence, more difficult to launch message-specific replay attacks. 
\end{itemize}
To further reduce the in-threat period for \gls{csmrpl}, we propose that \gls{rpl} should be forced to send the first \gls{uc} message as soon as it finishes processing the first \gls{mc} message.
\squeezeupBY{2}
\subsection{Trades-offs for \gls{csmrpl}}
It is clear from the analysis of the power consumption patterns (Figures \ref{fig_PWR-CMAC-I}, \ref{fig_PWR-NR-I}, \ref{fig_PWR-CMAC-E}, and \ref{fig_PWR-NR-E}) that \gls{csmrpl} power consumption is slightly higher than \gls{umrpl} and \gls{psmrpl} when there is no attack. This is mainly due to the \gls{sc} recovery mechanism, which was triggered several times during the experiments, mainly when there is a lot of traffic (e.g., at the network initialization phase or when there is an active attack), due to the increased number of lost/corrupt control messages. All of the reported results include the overhead of running the \gls{sc} recovery mechanism.
%It is clear from the analysis of the power consumption patterns (Figures \ref{fig_PWR-CMAC-I}, \ref{fig_PWR-NR-I}, \ref{fig_PWR-CMAC-E}, and \ref{fig_PWR-NR-E}) that \gls{csmrpl} power consumption is slightly higher than \gls{umrpl} and \gls{psmrpl} when there is no attack. In addition, \gls{csmrpl} requires a proper recovery mechanism for the SC values/table. %These are the main trade-offs for better security than \gls{umrpl} and \gls{psmrpl}. 
%It is worth mentioning that the \gls{sc} recovery mechanism was triggered several times during the experiments, mainly when there is a lot of traffic (e.g., at the network initialization phase or when there is an active attack), due to the increased number of lost/corrupt control messages. All of the experiments results include the overhead of running the \gls{sc} recovery mechanism.
On the advantage side, the evaluations proved the superiority of \gls{csmrpl} at mitigating the investigated attacks with higher data \gls{pdr}, lower data packet latency, and lower average power consumption compared to the other secure modes when under attacks.
\squeezeupBY{1.5}
\section{Conclusion}\label{Conc}
In this paper, we proposed a novel secure mode for \gls{rpl}, the \acrlong{csmrpl}, that is based on the concept of intra-flow \gls{nc}, to enhance \gls{rpl} security and to build a mitigation capability of replay attacks into the protocol itself, without significantly changing the way \gls{rpl} works. A prototype of \gls{csmrpl} was devised, and its security and performance were evaluated against the currently implemented secure modes of \gls{rpl} (\gls{umrpl}, \gls{psmrpl}, and \gls{psmrpl}rp) under three replay routing attacks (\gls{na}, \gls{ca}, and \gls{wh} attack). It was shown that \gls{csmrpl} successfully mitigated the replay attacks (\gls{na} and \gls{wh} attack) while significantly reduced the effect of \gls{ca}, all with latency and power consumption less than the other secure modes. Also, it was shown that \gls{csmrpl} has a significantly smaller in-threat period than all other secure modes. In addition, the ability to integrate external security mechanism opens the door to further enhance \gls{rpl} security through future expansions. For example, it is possible to support mobility vie suitable external security mechanism that would allow \gls{csmrpl} to trust the mobile nodes, which is left for future work.

\section*{Acknowledgment}
%(Acknowledgment will be added at the final submission)
The authors acknowledge support from the Natural Sciences and Engineering Research Council of Canada (NSERC) through the Discovery Grant program.
\squeezeup
%\newpage
\bibliographystyle{IEEEtran}
\bibliography{references}
%\newpage
\vskip -2\baselineskip plus -1fil
\begin{IEEEbiography}[{\includegraphics[width=1in,height=1.25in,clip,keepaspectratio]{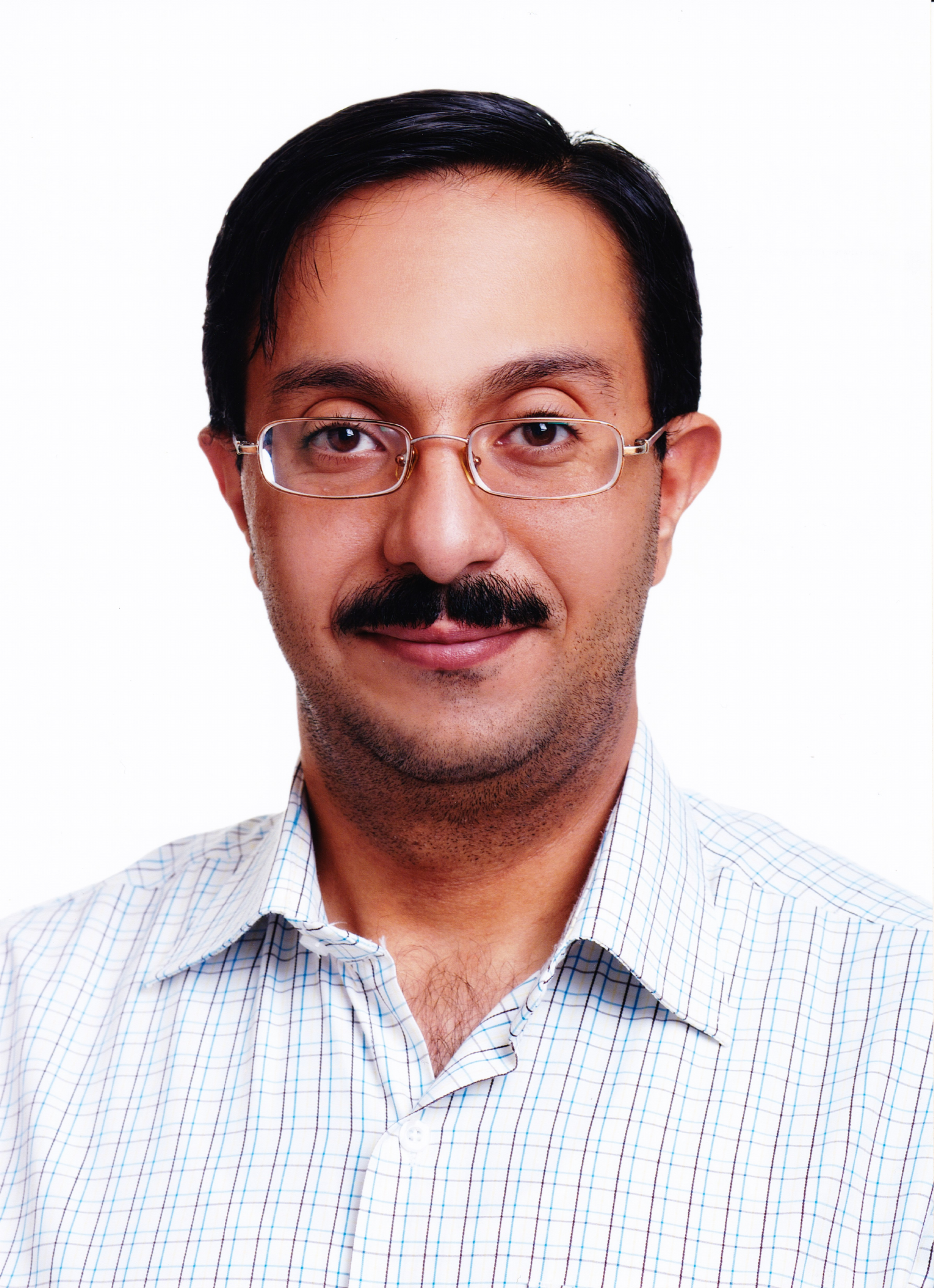}}]{Ahmed Raoof}
	%(Biography will be added at the final submission)
	received his Ph.D. degree (2021) from Department of Systems and Computer Eng., Carleton University, Ottawa, Canada, and both his M.Sc. (in Telecomm. Eng. - 2009) and B.Sc. (in Electrical and Electronic Eng. - 2003) degrees  from the University of Benghazi, Libya. He worked at the University of Benghazi as a lecturer in the Department of Comp. Net., Faculty of Information Technology (2009 - 2013). Currently, he is a postdoctoral fellow with the Department of Sys. and Comp. Eng. at Carleton University. His research focuses on the security of data networks and Internet of Things.
\end{IEEEbiography}
\vskip -2\baselineskip plus -1fil
\begin{IEEEbiography}[{\includegraphics[width=1in,height=1.25in,clip,keepaspectratio]{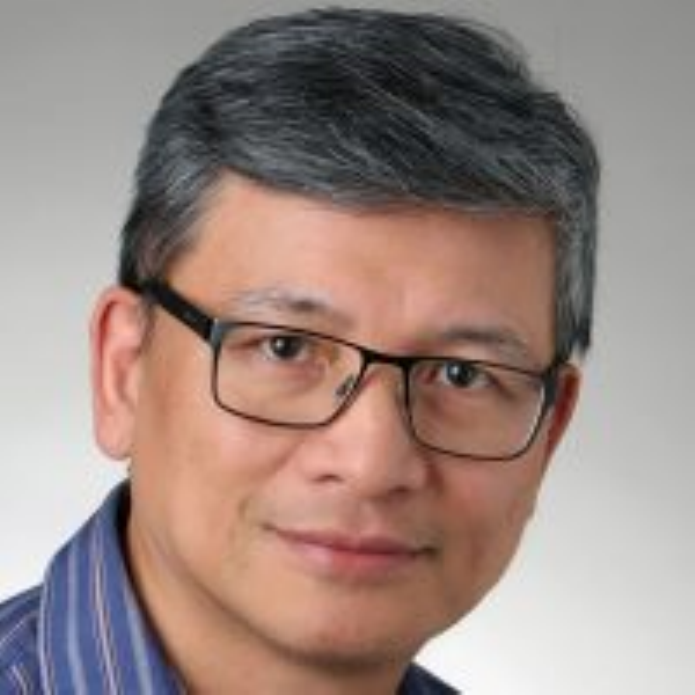}}]{Chung-Horng Lung}
	%(Biography will be added at the final submission)
	(Senior Member, IEEE) received the B.S. degree (1982) in Comp. Science and Eng. from Chung-Yuan Christian University, Taoyuan, Taiwan, and the M.S. (1988) and Ph.D. (1994) degrees in Computer Science and Eng. from Arizona State University ,Tempe, AZ, USA. He was with Nortel Networks, Ottawa, ON, Canada from 1995 to 2001. In September 2001, he joined the Department of Systems and Computer Eng., Carleton University, Ottawa, where he is currently a Professor. His research interests include: Communication Networks, Software Engineering, and Cloud Computing.
\end{IEEEbiography}
\vskip -2\baselineskip plus -1fil
\begin{IEEEbiography}[{\includegraphics[width=1in,height=1.25in,clip,keepaspectratio]{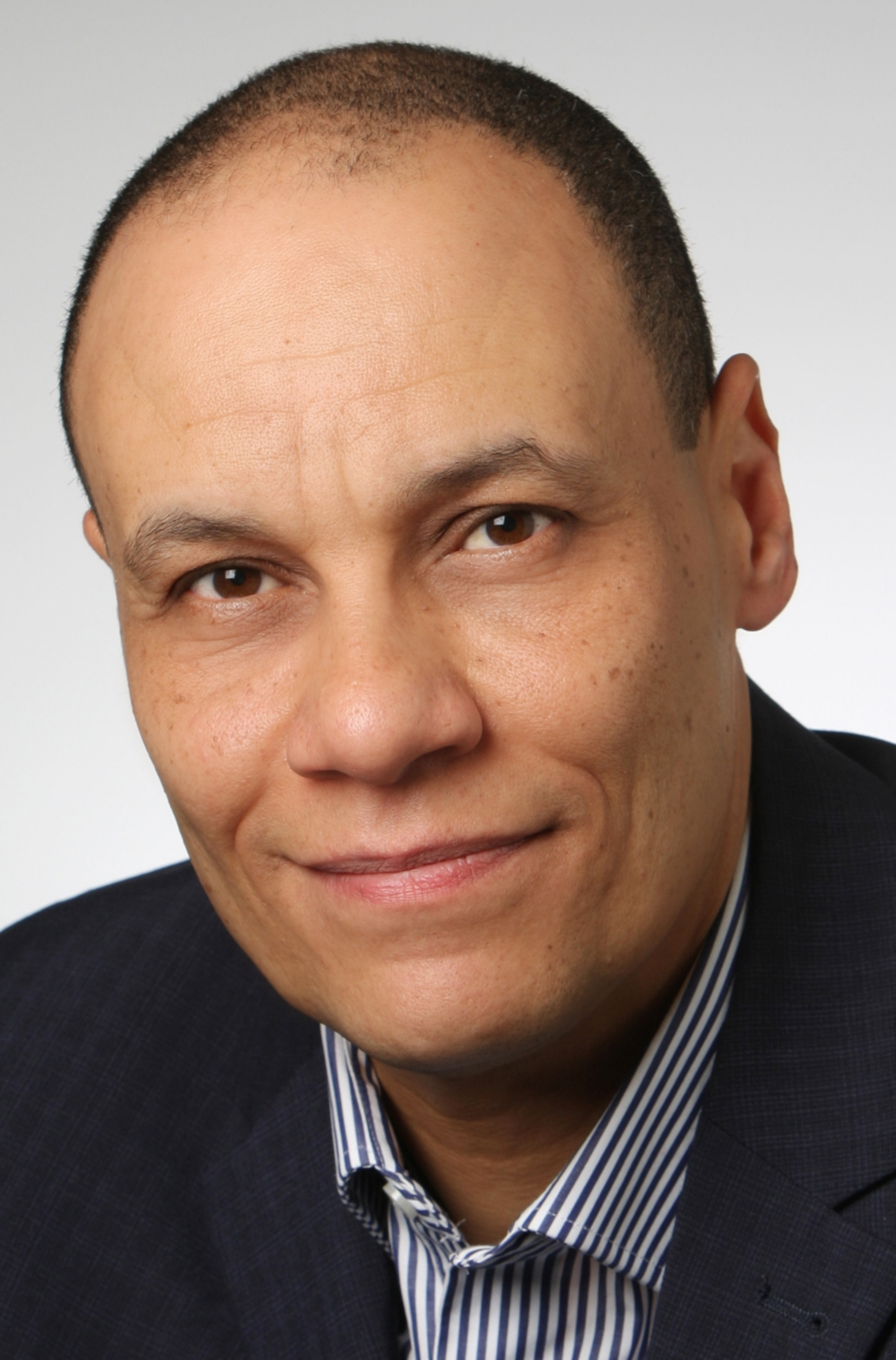}}]{Ashraf Matrawy}
	%(Biography will be added at the final submission)	
	(Senior Member, IEEE) is a Professor with the School of Information Technology, Carleton University, Ottawa, ON, Canada. He received his B.S. (1993) and M.S. (1998) degrees from Alexandria University, Alexandria, Egypt, and his Ph.D. (2003) from Carleton University, Ottawa, ON, Canada. He serves on the editorial board of the IEEE Commun. Surveys and Tutorials journal, and served as a technical program committee member of several IEEE conferences (CNS, ICC, Globecom, LCN) and IEEE/ACM CCGRID. He is also a Network Co-Investigator of Smart Cybersecurity Network (SERENE-RISC). 
\end{IEEEbiography}
%\vskip -2\baselineskip plus -1fil

\end{document}